\pdfoutput=1

\RequirePackage{fix-cm}
\documentclass[twocolumn]{svjour3}          

\smartqed  
\usepackage{graphicx}

\usepackage{comment}

\usepackage{amsmath}
\usepackage{amssymb}

\usepackage{datetime2}

\usepackage[misc,geometry]{ifsym} 

\usepackage[colorlinks=true, citecolor=blue, linkcolor=blue, urlcolor=blue, pdfborder={0 0 0}, breaklinks=true]{hyperref}
\makeatletter
\g@addto@macro{\UrlBreaks}{\do\-}
\g@addto@macro{\UrlBreaks}{\do\#}
\makeatother

\usepackage{cite}

\usepackage[switch]{lineno}
\newcommand*\patchAmsMathEnvironmentForLineno[1]{
  \expandafter\let\csname old#1\expandafter\endcsname\csname #1\endcsname
  \expandafter\let\csname oldend#1\expandafter\endcsname\csname end#1\endcsname
  \renewenvironment{#1}
     {\linenomath\csname old#1\endcsname}
     {\csname oldend#1\endcsname\endlinenomath}}
\newcommand*\patchBothAmsMathEnvironmentsForLineno[1]{
  \patchAmsMathEnvironmentForLineno{#1}
  \patchAmsMathEnvironmentForLineno{#1*}}
\AtBeginDocument{
\patchBothAmsMathEnvironmentsForLineno{equation}
\patchBothAmsMathEnvironmentsForLineno{align}
\patchBothAmsMathEnvironmentsForLineno{flalign}
\patchBothAmsMathEnvironmentsForLineno{alignat}
\patchBothAmsMathEnvironmentsForLineno{gather}
\patchBothAmsMathEnvironmentsForLineno{multline}
}

\newcommand{\doi}[1]{\url{https://doi.org/#1}}
\newcommand{\arxiv}[1]{\url{https://arxiv.org/abs/#1}}

\newcommand*{\affaddr}[1]{#1}
\newcommand*{\affmark}[1][*]{\textsuperscript{#1}}

\journalname{Computing and Software for Big Science}

\def\makeheadbox{{
\hbox to0pt{\vbox{\baselineskip=10dd\hbox
to\hsize{\kern3pt\vbox{\kern3pt
\hbox{\bfseries\hspace*{11.4cm}CERN-LPCC-2020-002}
\hbox{\bfseries\hspace*{11.4cm}FERMILAB-PUB-20-183-SCD-T}
\hbox{\bfseries\hspace*{11.4cm}MCNET-20-15}
\kern3pt}\hfil\kern3pt}}
}}}

\begin{document}

\title{Challenges in Monte Carlo 
event generator software\\ 
for High-Luminosity LHC}

\author{
The~HSF~Physics~Event~Generator~WG
\and
Andrea~Valassi~(editor)\protect\affmark[1] 
\and
Efe~Yazgan~(editor)\affmark[2] 
\and
Josh~McFayden~(editor)\affmark[1,3]
\and
Simone~Amoroso\affmark[4]
\and
Joshua~Bendavid\affmark[1]
\and
Andy~Buckley\affmark[5]
\and
Matteo~Cacciari\affmark[6,7]
\and
Taylor~Childers\affmark[8]
\and
Vitaliano~Ciulli\affmark[9]
\and
Rikkert~Frederix\affmark[10]
\and
Stefano~Frixione\affmark[11]
\and
Francesco~Giuli\affmark[12]
\and
Alexander~Grohsjean\affmark[4]
\and
Christian~G\"utschow\affmark[13]
\and
Stefan~H\"oche\affmark[14]
\and
Walter~Hopkins\affmark[8]
\and
Philip~Ilten\affmark[15,16]
\and
Dmitri~Konstantinov\affmark[17]
\and
Frank~Krauss\affmark[18]
\and
Qiang~Li\affmark[19]
\and
Leif~L\"{o}nnblad\affmark[10]
\and
Fabio~Maltoni\affmark[20,21]
\and
Michelangelo~Mangano\affmark[1]
\and
Zach~Marshall\affmark[3]
\and
Olivier~Mattelaer\affmark[21]
\and
Javier~Fernandez~Menendez\affmark[22]
\and
Stephen~Mrenna\affmark[14]
\and
Servesh~Muralidharan\affmark[1,8]
\and
Tobias~Neumann\affmark[13,23]
\and
Simon~Pl\"atzer\affmark[24]
\and
Stefan~Prestel\affmark[10]
\and
Stefan~Roiser\affmark[1]
\and
Marek~Sch\"onherr\affmark[18]
\and
Holger~Schulz\affmark[16]
\and
Markus~Schulz\affmark[1]
\and
Elizabeth~Sexton-Kennedy\affmark[14]
\and
Frank~Siegert\affmark[25]
\and
Andrzej~Si\'odmok\affmark[26]
\and
Graeme~A.~Stewart\affmark[1]
}
\authorrunning{The HSF Physics Event Generator WG}

\institute{
\Letter\ Andrea Valassi\\
andrea.valassi@cern.ch\\
\Letter\ Efe Yazgan\\
efe.yazgan@cern.ch\\
\Letter\ Josh McFayden\\
mcfayden@cern.ch\\
\\
\affaddr{\affmark[1]CERN, Geneva, Switzerland}\\
\affaddr{\affmark[2]National Taiwan University, Taipei, Taiwan}\\
\affaddr{\affmark[3]Lawrence Berkeley National Laboratory, USA}\\
\affaddr{\affmark[4]DESY, Hamburg, Germany}\\
\affaddr{\affmark[5]University of Glasgow, UK}\\
\affaddr{\affmark[6]LPTHE, Sorbonne Universit\'e and CNRS, Paris, France}\\
\affaddr{\affmark[7]Universit\'e de Paris, France}\\
\affaddr{\affmark[8]Argonne National Laboratory, USA}\\
\affaddr{\affmark[9]INFN and Universit\`a di Firenze, Italy}\\
\affaddr{\affmark[10]Lund University, Sweden}\\
\affaddr{\affmark[11]INFN and Universit\`a di Genova, Italy}\\
\affaddr{\affmark[12]INFN and Universit\`a Tor Vergata, Roma, Italy}\\
\affaddr{\affmark[13]University College London, UK}\\
\affaddr{\affmark[14]Fermi National Accelerator Laboratory, USA}\\
\affaddr{\affmark[15]University of Birmingham, UK}\\
\affaddr{\affmark[16]University of Cincinnati, USA}\\
\affaddr{\affmark[17]NRC ``Kurchatov Institute'' - IHEP, Protvino, Russia}\\
\affaddr{\affmark[18]University of Durham, UK}\\
\affaddr{\affmark[19]Peking University, Beijing, China}\\
\affaddr{\affmark[20]Universit\`a di Bologna, Italy}\\
\affaddr{\affmark[21]Universit\'e Catholique de Louvain, Belgium}\\
\affaddr{\affmark[22]Universidad de Oviedo, Spain}\\
\affaddr{\affmark[23]Illinois Institute of Technology, Chicago, USA}\\
\affaddr{\affmark[24]University of Vienna, Austria}\\
\affaddr{\affmark[25]Technische Universit\"at, Dresden, Germany}\\
\affaddr{\affmark[26]IFJ PAN and Jagiellonian University, Krakow, Poland}
}

\date{Version 4.0 (18 February 2021)}

\maketitle

\begin{abstract}
We review 
the main software and computing challenges 
for the Monte Carlo physics event generators
used by the LHC experiments,
in view of the High-Luminosity LHC (HL-LHC) physics programme.
This paper has been prepared 
by the HEP Software Foundation (HSF)
Physics Event Generator Working Group
as an input to the 
LHCC review of HL-LHC computing,
which has started in May 2020.
\end{abstract}

\section{Introduction}
\label{intro}
Physics event generators 
are one of the computational pillars of 
any High Energy Physics (HEP) experiment,
and in particular of 
the Large Hadron Collider (LHC) experiments.
In this paper,
we review the main software and computing 
challenges for the physics event generators  
used  by  the  
ATLAS~\cite{bib:atlas}
and
CMS~\cite{bib:cms} 
experiments,
in view of the 
high-luminosity running phase
of the LHC experimental programme (HL-LHC),
which should be operational from the end 
of 2027~\cite{bib:hllhc}.
This document has been prepared by
the Physics Event Generators
Working Group (WG)~\cite{bib:hsfgen2019}
of the HEP Software Foundation (HSF),
as an input to the 
review of the~HL-LHC computing strategy
by the LHC Experiments Committee
(LHCC)~\cite{bib:lhcc},
which has started in May 2020~\cite{bib:lhccreport1},
as previously 
planned~\cite{bib:lhcc136,bib:lhcc139,bib:lhcc140,bib:lhcc141}.
As is the case for the LHCC review,
this paper focuses 
on ATLAS and CMS,
but it also contains 
important considerations for
ALICE~\cite{bib:alice}
and LHCb~\cite{bib:lhcb}.
The HSF has also prepared a more general 
document~\cite{bib:hsflhcc} for the LHCC, 
which covers the status and challenges
in~the broader area 
of common tools and community~software.

This paper gives an overview 
of the many 
challenges in the generator area,
and of the work that can be done to address them. 
Its outline
is the following.
Section~\ref{sec:hsfgen} gives an overview
of the role and challenges 
of physics event generators in LHC computing,
summarising the steps which led
to the creation of the HSF generator WG,
and its current activities.
Section~\ref{sec:nontech}
describes the collaborative challenges
in the development, use and maintenance
of generator software 
for LHC physics.
Section~\ref{sec:tech}
gives more details about 
the computational anatomy
of physics event generators,
and the technical challenges
in their development 
and performance optimization.
Section~\ref{sec:phys}
summarizes some of the 
main open questions about 
the required physics accuracy of 
event generators at HL-LHC,
and their impact on computational costs.
Finally, in Sec.~\ref{sec:concl}
we compile a list
of high-priority items on which 
we propose that
the R\&D~on the computational 
aspects of generators
(and in particular the activities 
of the HSF generator WG)
should focus,
in view of the more in-depth 
LHCC review of HL-LHC software
that is currently scheduled 
for Q3 2021~\cite{bib:lhcc141}.

It should be stressed 
that this paper focuses 
on the software and computing aspects
of event generators,~rather than on 
the underlying physics.
To be able to describe
the overall computational goals and structure 
of these software applications
and put them in context,
many of the relevant physics concepts
are in any case mentioned
and briefly explained.
This is done
using a language that tries to be somewhat accessible
also to software engineers and computing experts
with no background in particle physics,
even if the resulting text is not meant
to be an exhaustive overview
of these complex issues
from a theoretical point of view.

One should also note that,
to some extent,
some of the issues described in this paper,
such as the collaborative challenges 
and human resource concerns
related to the development
and support of generator software,
have already been raised
in previous community efforts.
These include, in particular,
the HSF Community White Paper 
(CWP)~\cite{bib:hsfcwp}
and the document~\cite{bib:buckley2019mcnet}
that was submitted as an input
to the Open Symposium~\cite{bib:espp}
on the Update of European 
Strategy for Particle Physics.

\section{The HSF Physics Event Generator WG}
\label{sec:hsfgen}

\sloppy
Physics event generators 
are an essential 
component of the data processing and analysis chain 
of the LHC experiments,
and a large consumer of 
resources in the Worldwide LHC Computing Grid 
(WLCG)~\cite{bib:wlcg}.
All of the scientific results
of the LHC experiments,
such as 
precision measurements
of physics parameters and searches for new physics, 
depend significantly 
on the comparison of experimental measurements 
to theoretical predictions,
in most cases
computed using \mbox{generator~software}. 

\sloppy
Using Monte Carlo (MC) techniques, 
generators allow both the calculation
of differential and total cross sections
and the generation of weighted 
or unweighted events 
for experimental studies 
(this is explained in more detail 
in Sec.~\ref{sec:anatomy},
where these concepts are briefly defined).
Within the experiments, 
generators are used primarily 
to produce large samples 
of (mostly unweighted) events:
this is the first step in the production chain 
for simulating LHC collisions, 
which is followed by 
detector simulation and event reconstruction.
In each of the two general purpose 
LHC experiments, ATLAS and CMS, 
the overall number of events that are 
generated by the central production teams 
and passed through full detector simulation 
and event reconstruction, 
across all relevant physics processes, 
is of the order of magnitude of O($10^{10}$) 
events for every year of LHC data taking. 
Typically, the sizes of these samples 
of simulated events are approximately 
a factor of 3 larger than the overall 
number of data events collected 
during the corresponding time range.
These large-scale event generation campaigns 
have a computational cost,
mainly in terms of the ``compute'' (i.e. CPU) resources used,
the majority of which are provided by the
WLCG infrastructure.
The limited size of the simulated samples 
that can be produced under resource constraints
is a source 
of major uncertainty in many analyses
(for example, in Higgs boson measurements
of both ATLAS~\cite{bib:hatlas}
and CMS~\cite{bib:hcms}).
This is an issue 
which is limiting the potential 
physics output of the LHC programme, 
and may get significantly worse at HL-LHC, 
where the projected computing needs of the experiments
exceed the resources that are expected
to be available~\cite{bib:hsfcwp}, 
despite the fact that 
the most aggressive HL-LHC physics 
projections~\cite{bib:yellow1,bib:yellow2,bib:yellow3,bib:yellow4}
assume no uncertainty due to the 
limited size of simulated samples.

\fussy
When the HEP Software Foundation
prepared its CWP~\cite{bib:hsfcwp} in 2017,
the fraction of the ATLAS CPU resources 
in WLCG used for event generation 
was estimated~\cite{bib:hsfgen2017} at around 20\%.
Beyond the existing projections, 
which assume the same level 
of theoretical precision
as in the current event generation campaigns, 
concern was also raised that 
event generation would become computationally 
more expensive at the HL-LHC, 
where more complex calculations 
(e.g. beyond next-to-leading-order
or with higher jet multiplicities) 
will be needed~\cite{bib:maltoni}. 
It was thus clear that
speedups in generator software are 
needed to address the overall 
computing
resource problem expected at the HL-LHC.
This is of course also the case 
for the other big consumers of CPU 
(detector simulation and reconstruction),
but until now these areas have had more focus,
and significant speedups are already expected 
on the HL-LHC timescales,
which has not been the case for generators.
Other issues in the generator area, 
both technical and non-technical 
(e.g. funding, training and careers)
also became obvious 
while preparing the CWP.

For these reasons, 
the HSF organised a three-day 
Workshop~\cite{bib:hsfgen2018,bib:buckley2019acat} 
at the end~of 2018 to focus on the 
software and computing aspects 
of event generators. 
Their usage in the experiments 
was reviewed,
revealing a large discrepancy 
in the CPU budgets quoted
by ATLAS and CMS, 14\% and 1\%, 
respectively, for 2017~\cite{bib:sexton2018}. 
This was attributed,
at least partly, to the different packages 
and parameter settings used
by the two experiments,
but it was clear that further studies were needed.

\fussy
A Working Group of the HSF 
on Physics Event Generators~\cite{bib:hsfgen2019}
was therefore set up at the beginning of 2019. 
The main focus of the WG so far has been 
to get a better understanding 
of the current usage of generators in the experiments, 
and to identify and prioritise 
the areas where computing costs can be reduced. 
In particular, the ATLAS and CMS compute budgets 
have been analysed in detail:
currently, it is estimated 
that the fractions of WLCG compute allocations 
used for generation today are around 
12\% for ATLAS and~5\% for CMS. 
In terms of absolute CPU time
spent for event generation,
the ratio between ATLAS and CMS 
is actually larger, 
as the overall ATLAS budget 
for compute resources is larger than that of CMS. 
To understand what causes this difference,
detailed benchmarking
of the computational costs of
Sherpa~\cite{bib:sherpa}
and MadGraph5\_aMC@NLO~\cite{bib:mg5amc}
(in the following abbreviated as MG5\_aMC)
have also started~\cite{bib:josh2018,bib:efe2018},
as these are the two generators
used for some of the most expensive 
event generation productions
in ATLAS and CMS, respectively.
The WG has also been active in other areas, 
such as in discussing the possible sharing 
of common parton-level samples 
by ATLAS and CMS~\cite{bib:javier19}, 
and in reviewing and supporting 
the efforts for porting generators 
to modern architectures, notably GPUs. 
This last activity is particularly important, 
as it has become increasingly clear 
that being able to run compute-intensive 
WLCG software workloads 
on GPUs~\cite{bib:avgpus}
would allow the exploitation 
of modern GPU-based supercomputers 
at High Performance Computing (HPC) 
centers, 
and generators look like 
a natural candidate for this,
as discussed later on in Sec.~\ref{sec:gpus}.

\sloppy
Looking forward,
the WG plans to continue its activities 
in the areas described above,
but also to expand it in
a few other directions.
One of the goals of this paper
is that of dissecting and analysing
the many different challenges,
both technical and non-technical, 
in the generator domain,
to identify the specific areas
where work is most urgently needed,
or where the largest improvements
are expected to be possible to reduce 
the gap between required and available 
computing resources at the time of HL-LHC.
It should also be pointed out
that the role of the WG in this context
is mainly that of providing a forum
for information exchange,
and possibly supporting and coordinating
common activities involving
the collaboration of several teams
or the comparison of their results,
but most of the concrete work
is generally expected to be done 
by the individual experiments
or theoretical \mbox{physicist teams}.

\fussy
\section{Collaborative challenges}
\label{sec:nontech}

In this section,
we give an overview 
of the collaboration challenges
in the development, use and maintenance
of generator software for LHC.
By and large,
these are mainly non-technical challenges
that concern human resources,
i.e. actual people,
and their organisation,
training and motivation,
rather than computing resources,
software modules
or theoretical physics models.

\subsection{A very diverse software landscape}
\sloppy
The landscape of generator software 
is extremely varied,
even more than 
in detector simulation, 
event reconstruction 
or analysis workloads. 
For a review, see for instance 
Refs.~\cite{bib:buckley2011,bib:sjostrand2012,bib:sjostrand2016,bib:buckley2019mcnet}.
Different generators 
(Sherpa, MG5\_aMC, 
the POWHEG BOX~\cite{bib:POWHEGbox}, 
Pythia~\cite{bib:pythia82}, 
Herwig~\cite{bib:herwigpp,bib:herwig70,bib:herwig71}, 
Alpgen~\cite{bib:alpgen}, etc.)
are used in the community,
mainly for two reasons: 
firstly, one needs multiple 
independent calculations with 
potentially different approximations 
to cross-check one another; 
and secondly, the different generators 
vary in their features (for example, 
some might simulate only a subset 
of the physics processes of interest). 
A given process may be simulated
with a different physics precision,
e.g. leading-order (LO),
next-to-leading-order (NLO), 
or next-to-next-to-leading-order (NNLO)
in a power series expansion 
in the strong-force ``coupling constant''.
Generating a sample also involves 
choices of 
hadronization and parton shower (PS) models 
(Pythia~\cite{bib:pythia82}, 
Herwig~\cite{bib:herwigpp,bib:herwig70,bib:herwig71}, 
Ariadne~\cite{bib:ariadne}, etc.), 
underlying event 
tunes~\cite{bib:atlastune,bib:cmstune,bib:newtunes,bib:autotune}, 
prescriptions for 
matching/merging\footnote{
\hspace*{-1mm}
In this paper, we use 
the definitions of matching and merging 
given in Ref.~\cite{bib:fxfx},
which are briefly hinted at 
in Sec.~\ref{sec:inefficiencies}.}
(MC@NLO~\cite{bib:mcnlo}, 
POWHEG~\cite{bib:POWHEG}, 
KrkNLO~\cite{bib:krknlo}, 
CKKW~\cite{bib:ckkw}, 
\mbox{CKKW-L}~\cite{bib:ckkwl}, 
MLM~\cite{bib:mlm,bib:alwall08}, 
MEPS@NLO~\cite{bib:mepsnlo}, 
MINLO~\cite{bib:minlo}, 
FxFx~\cite{bib:fxfx}, 
UNLOPS~\cite{bib:unlops},
Herwig7 Matchbox~\cite{bib:platzer2012,bib:platzer2013,bib:bellm2018},
etc.),
\mbox{``afterburner''} tools 
for simulating particle decays 
and quantum electrodynamics (QED) 
radiative corrections
(\mbox{EvtGen}~\cite{bib:evtgen}, 
Tauola~\cite{bib:tauola}, 
Photos~\cite{bib:photos}, etc.),
and other input parameters such as
parton distribution functions 
(PDFs)~\cite{bib:pdf4lhc},
primarily via the LHAPDF
library~\cite{bib:lhapdf}.

Various combinations of software libraries 
are thus possible, 
often written by different authors 
and some dating back many years,
reflecting theoretical research
within different teams.
For a given process, 
the LHC experiments often use different 
software packages and settings from one another, 
and a single experiment can generate events 
using more than one choice. 
Many different packages and configurations 
may therefore need to be studied and improved 
to get cumulative CPU cost reductions. 
The large number of packages 
also complicates their long-term maintenance
and integration in the experiments 
software and workflows, 
sometimes leading to Grid job failures 
and computing inefficiencies.
Other packages are also absolutely critical 
for the whole generator community
and must be maintained, 
even if their CPU cost is relatively low 
(Rivet~\cite{bib:rivet}, 
Professor~\cite{bib:professor},
HepMC~\cite{bib:hepmc,bib:hepmc3}, 
FastJet~\cite{bib:fastjet},~etc.).

\fussy
\subsection{A very diverse human environment}

\sloppy
A broad spectrum
of skills and profiles are needed
for the development and support of event generators:
theorists (who create fundamental physics models, 
and design, develop and optimize most generator code); 
experimentalists working on research 
(who determine which types 
of event samples are required,
and of which size);
experimentalists working on computing 
(who implement, monitor and account 
execution of workflows on computing resources); 
software engineers and system performance experts 
(who may help to analyse and improve 
the efficiency of software applications 
and deployment models). 
This is a richness and opportunity, 
as some technical problems 
are best addressed by people 
with specific skills, 
but it also poses some challenges
as these technical problems are best addressed
by bringing all these people together.
Facilitating this cross-collaboration 
is one of the main goals of the WG.

\fussy
\paragraph{Training challenges.}
Theorists and experimentalists 
often lack formal training 
in software development and optimization. 
Software engineers, 
but also many experimentalists, 
are not experts 
in the theoretical physics models 
implemented in MC codes.

\sloppy
\paragraph{Communication challenges.}
It is difficult to find a shared 
terminology and set of concepts 
to understand one another: 
notions and practices 
that are taken for granted in one domain 
may be obscure for others. 
An example: there are many articles 
about the physics in generators, 
but software engineers 
would need papers describing 
the main software modules 
and overall data and control flow.
Similarly, there are only very 
few articles where the experiments
describe the software and computing 
workflows of their large scale MC productions
(Ref.~\cite{bib:roiser} is one such
example for LHCb).

\fussy
\paragraph{Career challenges.} 
Those working in the development, 
optimization and execution 
of generator software 
provide essential contributions 
to the success of the 
HL-LHC physics programme 
and it is critical that they get 
the right recognition and motivation. 
However, theorists, in general, get recognition 
from the papers they publish
and from the citations on these,
and they may not be motivated 
to work on software optimizations 
that do not have enough 
theoretical physics content 
to advance their careers. 
Generator support tasks 
in the experiments may also 
not be valued enough to secure jobs
or funding to experimentalists 
pursuing a career in~research.

\paragraph{Mismatch in usage patterns and in optimization focus.} 
The way generators are built 
and used by their authors 
is often different from the way 
in which they are deployed 
and integrated by the experiments 
in their software frameworks 
and computing infrastructure.
The goals and metrics 
of software optimization work 
may also differ,
as discussed more in detail
in Sec.~\ref{sec:tech}.
Theorists, 
who typically work with weighted events
and fast detector parametrizations if any,
are mainly interested 
in calculating cross sections 
and focus on minimising 
the phase space integration time 
for a given statistical precision. 
The LHC experiments typically run 
large scale productions for
generating fully exclusive events,
which are mostly unweighted
as they must be processed 
through expensive detector simulation 
and event reconstruction steps:
therefore, they need to maximize 
the throughput of events generated 
per unit time on a given computing system.

\paragraph{Programming languages.}
Attracting collaborators 
with a computer science background 
to work on generators, especially students, 
may also be complicated by the fact 
that critical components 
of some generator packages 
are written in Fortran,
which is rarely used in industry
and less popular among developers
than other programming languages.
Some of the generators 
also do not use industry standard 
version control systems, 
making it harder to contribute code.

\section{Technical challenges}
\label{sec:tech}

In this section, 
we give more details about 
the technical challenges 
in the software development 
and performance optimization
of MC physics event generator codes.
To this end,
it is useful to first give 
a brief, high-level, reminder 
of their computational goals 
and internal data flows,
and of the typical production workflows
used by the experiments.

\subsection{Computational anatomy of a MC event generator}
\label{sec:anatomy}
\newcommand{\xvec}{\mathbf{x}}
\newcommand{\fxvec}{f(\xvec)}
\newcommand{\gxvec}{g(\xvec)}
\newcommand{\wxvec}{w(\xvec)}
\newcommand{\xveci}{{\xvec_i}}
\newcommand{\fxveci}{f(\xveci)}
\newcommand{\gxveci}{g(\xveci)}
\newcommand{\wxveci}{w(\xveci)}
\newcommand{\wi}{{w_i}}
\newcommand{\wmax}{w_\text{max}}
\newcommand{\njets}{n_\text{jets}}
\newcommand{\xs}{\sigma}
\newcommand{\lumi}{{\mathcal{L}}}
\newcommand{\nexp}{{N_\mathrm{exp}}}
\newcommand{\dxs}{d\xs}
\newcommand{\dxvec}{d\xvec}
\newcommand{\dxvecn}{d\xvecn}
\newcommand{\dxsdxvecn}{\frac{\dxs}{\dxvecn}}
\newcommand{\accO}{{\Omega_O}}
\newcommand{\dO}{dO}
\newcommand{\dxsdO}{\frac{\dxs}{\dO}}
\newcommand{\pdf}{p}
\newcommand{\xveco}{x_{1}}
\newcommand{\xvect}{x_{2}}
\newcommand{\xvecn}{\Phi_{\!n}}
\newcommand{\xvecnp}{\Phi_{\!n+1}}
Particle physics
is based on quantum mechanics,
whose description of Nature 
is intrinsically probabilistic.
The predictions of HEP theoretical models
that are numerically computed 
in event generators
(through a combination of
quantum field theory methodologies
and phenomenological approximations),
and which can be compared to experimental
measurements,
ultimately consist of probabilities 
and probability density functions.

In particular,
the probability that 
a collision ``event''
with a given ``final state'',
i.e. including $n$ particles 
of given types,
is observed in the collision
of the LHC proton beams,
is expressed in HEP 
in terms of the concept 
of a ``cross section''.
In general terms,
a cross section~$\xs$ represents 
the number of events $\nexp\!=\!\xs\lumi$
that are expected
per unit~``integrated luminosity'' 
$\lumi$ of the colliding beams 
(a~parameter that depends 
on their intensities and geometries,
and on the overall duration
of data-taking time).
More in detail,
a differential cross section, $\dxsdO$,
with respect to an observable $O$ 
(such as a rapidity 
or a transverse momentum), 
refers to the observation
of the desired final state
at different points~$\dO$
of the observable ``phase space'';
conversely, 
its integral 
$\sigma\!=\!\int_\accO\!\dxsdO\dO$
is referred to as the total cross section, 
if over the entire phase space,
or as a fiducial cross section, 
if over a well delimited region
$\accO$ of the phase space
(the so-called acceptance).

\sloppy
In this context,
the computational core 
of a physics event generator 
is the code that 
numerically calculates,
from first principles,
the fully differential cross section
$\dxsdxvecn\!(\xvec)$ 
for the ``hard scattering'' process
that leads to the desired
$n$-particle final state;
this is computed
as a function of the complete
kinematical configuration $\xvec$ 
of the elementary particles,
or ``partons'', involved
in this ``hard interaction''
for an individual collision event.
In the majority of cases, 
the calculation of $\dxsdxvecn$
is implemented by identifying 
all Feynman diagrams contributing 
to this process,
and calculating the ``invariant amplitude'' 
or ``matrix element'' (ME) 
for all of these diagrams combined
(although there are also generators
where matrix elements are computed
using algorithms not based on Feynman 
\mbox{diagrams~\cite{bib:alpgen,bib:caravaglios})}.

\fussy
For LHC processes,
the kinematical configuration
$\xvec\!=\!\{\xveco,\xvect,\xvecn\}$ 
of a collision event
essentially consists of 
a vector $\xvecn$,
including four real numbers
(related to their energy, mass and directions)
for each of the $n$ outgoing (final state) partons,
and of 
two real numbers $\xveco$ and $\xvect$
representing the momentum fractions 
of the two incoming (initial state) partons.
As described later on
in Eq.~\ref{eq:convolution},
$\dxsdxvecn\!(\xvec)$ is,
together with two 
parton distribution functions 
$\pdf(\xveco)$ and $\pdf(\xvect)$,
the central ingredient 
in the computation of a function $\fxvec$,
which essentially describes
the probability distribution
in the space of all possible 
kinematical configurations $\xvec$,
and from whose integral in this space
other relevant cross sections
may be computed, 
$\xs\!=\!\int\!\fxvec\dxvec$.

\paragraph{Integration and unweighted event generation.}
Given
the function $\fxvec$,
physics event generators 
are commonly used in HEP to 
solve two types 
of computational problems,
which are related to each other
and generally addressed
within a same execution of the software,
as discussed more in detail later on.
The first goal
(``phase space integration'') is 
to compute a cross section
as the integral of $\fxvec$ 
over the relevant phase space region.
The second goal
(``unweighted event generation'') is
to draw random samples of events 
whose kinematical configurations 
$\xvec$ are
distributed according to 
the theoretical prediction $\fxvec$.

\sloppy

Both of these goals
are achieved using 
Monte Carlo (MC) methods
(see Refs.~\cite{bib:james1968,bib:james1980}
for early reviews~of~this technique in HEP).
The distinctive feature of MC~methods 
is their reliance on 
the generation of random numbers
(or, more precisely, of 
``pseudo-random''~\cite{bib:rng} 
numbers)\footnote{
  As discussed in Ref.~\cite{bib:weinzierl2000},
  phase space integration may~also 
  be performed using classical numerical 
  methods (which do not belong to the MC category),
  or ``quasi-MC'' methods 
  based on ``quasi-random'' numbers~\cite{bib:james1980,bib:quasi}.
  The classical methods,
  such as Newton-Cotes formulas
  and Gaussian quadrature rules
  (one example is the Gauss-Kronrod algorithm
  which is used for numerical integration
  in the TOP++~\cite{bib:toppp} program),
  work well for one-dimensional 
  problems, 
  but tend to be inefficient
  for multi-dimensional integrals, where
  MC methods using pseudo-random numbers 
  converge much faster, 
  and quasi-MC methods using quasi-random numbers 
  even faster.
  Unlike phase space integration, however,
  unweighted event generation
  can only be addressed 
  by~MC methods using pseudo-random numbers,
  as neither of the other approaches is applicable:
  classical integration methods 
  because they do not involve random numbers,
  and quasi-MC techniques because 
  quasi-random numbers in a sample 
  are highly correlated to one another.
}.
In particular,
the starting point 
of both MC phase space integration
and MC unweighted event generation
is the calculation of $\fxvec$ 
for a large sample of events 
\mbox{$\xveci\!\in\!\{\xvec_1, \ldots, \xvec_N\}$},
drawn at random from~a~known 
probability density function $\gxvec$. 
More specifically:
\begin{enumerate}
\fussy
\item 
MC phase space integration
consists in drawing
a random sample of events $\xveci$
from the sampling function $\gxvec$,
and in numerically calculating 
an estimator of the integral 
$\xs\!=\!\int\!\fxvec\dxvec$,
as the average of the ``weight'' 
$\wi\!=\!\wxveci\!=\!\fxveci/\gxveci$ 
for all the events $\xveci$ in the sample.
It should be noted that this 
is not a deterministic approach,
in the sense that the result
of the calculation may change
if a different random sample is used:
it is easy to show, however,
that the estimator is unbiased,
and that its variance decreases as 1/$N$
if the number of events $N$
in the sample is increased.
From a software point of view,
the output of 
MC phase space integration\footnote{
  To avoid misunderstandings,
  it should be noted that,
  in an inconsistent way,
  the term ``phase space integration''
  is also commonly used
  to indicate the computational step
  in the software
  before unweighted event generation,
  which is needed 
  not only to compute a first coarse 
  estimate of cross sections, 
  but also to iteratively
  optimize the sampling algorithm
  (i.e. the choice of the 
  sampling function $\gxvec$),
  and to compute the maximum
  value $\wmax$ of $\wxvec$
  over the relevant region 
  of phase space.
  This is further discussed below.}
is essentially only one number, 
the estimate of the integral 
$\xs\!=\!\int\!\fxvec\dxvec$
over the acceptance $\accO$;
alternatively,
several numbers may also be calculated,
representing the values of $\dxsdO$ 
computed as the MC integrals of $\fxvec$
over different regions 
of phase space
within~$\accO$.

\item
MC unweighted event generation
consists in drawing
a random sample of events $\xveci$
from the sampling function $\gxvec$,
and in randomly rejecting 
some of them depending 
on the ratio of $\wxveci$ 
to the maximum weight $\wmax$ 
over the phase space.
For each event,
an accept-or-reject 
(or ``hit-or-miss'') 
decision is taken 
by drawing a random number $R$
uniformly distributed between 0 and 1:
the event is accepted 
if $R\!<\!\wxveci/\wmax$,
and rejected otherwise.
The resulting events,
whose distribution is now
described by $\fxvec$
rather than by $\gxvec$,
are referred to as ``unweighted'' 
in the sense that they~all 
have the same weight, 
which by convention is equal to~1. 
A special case of unweighting,
producing events whose weights
can be either ~+1 or~-1,
exists for calculations
leading to events with negative weights:
this is described later on.
From a software point of view,
the output of 
MC unweighted event generation
is a sample of events,
i.e. essentially 
a sample of vectors $\xveci$.
\end{enumerate}
\sloppy
The choice of the sampling algorithm 
(e.g. VEGAS~\cite{bib:vegas1978,bib:vegas1980}), 
or equivalently of the function $\gxvec$, 
is very important.
The closer $\gxvec$ is to $\fxvec$, 
that is to say the more constant 
the weight $\fxvec/\gxvec$ is 
over the entire phase space, 
the more precise is the integration 
(i.e. the lower the variance on the result) 
for a given sample size, 
and the more efficient 
is the unweighting procedure 
(i.e. the lower the fraction of events rejected).

It should be noted that
the experiments also do physics analysis 
with samples of weighted events,
which they produce for instance 
through ``biasing'' techniques,
as discussed in Sec.~\ref{sec:inefficiencies}.
Wherever possible, however,
unweighted events (and in particular
events with a positive weight~+1)
are preferred, as smaller event samples
are required than when using events
with non-uniform weights,
resulting in overall savings
of compute and storage resources.

\paragraph{Internal software workflow.}
Schematically, the internal software workflow 
of a typical generator 
is the following: 
first, 
when necessary
(i.e. when the process is too complex
to be manually hardcoded in advance),
the source code to compute 
the differential cross section 
$\dxsdxvecn$ of the hard process,
which is needed to derive $\fxvec$,
is produced through automatic code generation, 
after identifying  the relevant Feynman diagrams; 
a ``phase space integration'' step follows, 
where event samples are iteratively drawn 
not only to provide a first coarse estimate
of the relevant cross sections, but also
to optimize the sampling function $\gxvec$ 
and to estimate the maximum weight $\wmax$;
parton-level unweighted events 
are then generated using
the final, frozen, $\gxvec$ and $\wmax$;
parton showers, 
hadronization and
hadron decays to stable particles 
are finally applied 
on top of those ``parton-level'' events. 
During the unweighted event generation step,
``merging'' prescriptions
(described in more detail in Sec.~\ref{sec:inefficiencies})
may also need to be applied,
after parton showers 
and before hadronization;
experiment-level filters
and other techniques such as
forced decays or forced fragmentation
may also be applied~\cite{bib:ilten,bib:davis},
for instance to produce event samples
containing specific decays of~$B$~hadrons.

\fussy
The internal workflow of a generator application
is actually more complex
than described above,
because many different hard interactions
may contribute to the simulated process.
To start with,
for hadron colliders like the LHC,
the hard interactions take place
not between two protons,
but between two of the partons
in their internal substructure
(quarks of different flavors, and gluons):
this implies that separate integrals
for all possible 
types of
initial state partons,
using different sets of diagrams 
and of functions~$\fxvec$,
must be considered.
Using the factorisation 
theorem~\cite{bib:collins89},
which allows separating
perturbative (i.e. ME) 
and non-perturbative 
(parton distribution function) 
calculations
in quantum chromodynamics (QCD),
the total cross section
may be written~\cite{bib:buckley2011}~as
\begin{equation}
\xs \!=\!\!\! \sum_{a,b}\!
\int\!\! d\xveco \pdf_a(\xveco)\!\!
\int\!\! d\xvect \pdf_b(\xvect)\!\!
\int\!\! d\xvecn
\frac{\dxs_{\!ab}}{d\xvecn}(\xveco,\xvect,\xvecn), \!
\label{eq:convolution}
\end{equation}
i.e. as the convolution,
by the appropriate parton distribution functions
$\pdf_a(\xveco)$ and $\pdf_b(\xvect)$,
of the differential cross section
$\frac{\dxs_{ab}}{d\xvecn}$
for the production 
of $n$ final state particles
with properties $\xvecn$,
in the hard interaction 
of two partons of types $a$ and $b$
with momentum fractions $x_1$ and $x_2$,
respectively.

In addition, NLO calculations
imply the need to compute two separate
classes of integrals, which
involve two different classes
of Feynman diagrams 
and of functions~$\fxvec$, 
because matrix elements need to be separately
computed for standard ``$\mathbb{S}$-events'' 
and hard ``$\mathbb{H}$-events''~\cite{bib:mcnlo},
i.e. for final states with 
$n$ body kinematics $\xvecn$
(at tree level and one loop) 
and $n$+1 body kinematics $\xvecnp$
(at tree level),
respectively;
``matching'' prescriptions are then needed
to ensure that parton showers 
are used appropriately
in both types of events
(see also for instance 
Refs.\cite{bib:talk-weinzierl-2012,bib:talk-zaro-2015,bib:talk-luisoni-2019}
for detailed presentations that include
a graphical representation of these issues).
In NNLO~calculations,
the situation is similar 
to that of NLO calculations,
and even more complex.

\paragraph{Experiment production workflows.}
Phase space integration
(i.e. the optimization of the 
phase space sampling algorithm)
is a resource intensive step, 
but in many cases
it is only executed once 
in a given experiment production; this 
is known as the creation 
of ``gridpacks'' in MG5\_aMC and POWHEG, 
or ``sherpacks'' or ``integration grids'' 
in Sherpa\footnote{
  While ``gridpacks'' and ``integration grids''
  serve essentially the same purpose
  in different generators
  and contain similar information
  (the parametrization of an optimized
  and frozen tuning of the sampling algorithm,
  as well as an estimate of the maximum 
  event weight over the phase space),
  it should be noted that 
  the word ``grid'' 
  in these two terms alludes 
  to two very different concepts.
  The term 
  gridpacks~\cite{bib:gridpacks1,bib:gridpacks2},
  or ``Grid packages'' refers in MG5\_aMC to
  packages suitable to be sent over
  for event generation on Grid nodes
  (e.g. on the nodes provided by WLCG computing sites).
  The term ``integration grid'', conversely,
  refers to the partitioning of
  multi-dimensional phase space into hypercubes
  in the VEGAS adaptive sampling algorithm,
  which is used by default 
  in Sherpa~\cite{bib:vegas1980,bib:sherpagrid}.
}.
For instance,
creating a typical MG5\_aMC gridpack
for V+jets 
(i.e. a W or Z vector boson
produced in association with quarks or gluons) 
at NLO
may take up to several weeks 
on one multi-core node, 
or up to several days
in a typical cluster usage scenario;
see also Ref.~\cite{bib:talk-efe-2018}
for further details 
about how gridpacks are used in CMS. 
The generation of unweighted event samples, 
conversely, is where 
the LHC experiments spend essentially 
all of their 
yearly
generator CPU budgets: 
when pre-computed integration grids are available, 
this typically involves 
many Grid jobs submitted in parallel 
with different random number seeds
and thus unrelated to one another,
all of them reading 
the same integration grids as an input 
and storing events on their own output files. 
In principle, every Grid job could also 
go through the whole event generation chain, 
including both phase space integration 
and unweighted event generation, 
but this is an inefficient workflow 
which the experiments only use 
in specific cases, e.g.
for productions involving 
simple physics processes 
or few events,
where phase space integration 
is relatively fast and inexpensive and 
where the overhead from repeating it 
in each Grid job 
is negligible with respect 
to the overall CPU cost of the production,
or for generators lacking 
the option to create integration grids.
It should be noted, in any case, that also 
the workflows involving one gridpack creation 
and several unweighted event generation jobs 
can be somewhat inefficient, 
if the initialisation phase of each Grid job 
is not negligible with respect 
to its overall duration; 
this may happen, for example, 
if the pre-computed integration grids 
are very large and take a long time 
to load~\cite{bib:lange}.

\paragraph{Computational costs.}
The computational cost of a MC application
roughly scales with the number of points~$\xvec$ 
where the function $\fxvec$ is computed.
This is true both for gridpack creation,
where the cost scales with the~number of events
sampled during phase space integration
(itself a function of the accuracy
required for this step),
and for unweighted event generation,
where the cost scales with 
the overall number of events drawn 
prior to rejection by the unweighting algorithm.
As a consequence,
the most obvious approach
to reduce the overall computational
cost of event generation
is simply to try and decrease 
the number of points~$\xvec$ 
for which $\fxvec$ is computed.
This is described
in detail in Sec.~\ref{sec:inefficiencies},
where the possible reduction 
of many large inefficiencies 
in unweighted event generation is discussed,
as well as possible strategies 
for reusing events for more than one goal.

\sloppy
In addition, 
the intrinsic cost per event of computing $\fxvec$
approximately
scales itself with the number of Feynman diagrams
contributing to that process.
In particular, 
with respect to LO calculations 
for a given process,
NLO and especially NNLO calculations
for the same process
involve much higher numbers of diagrams,
some of which (``loop diagrams'') 
are also intrinsically more complex to compute.
Matrix element calculations are in fact performed
as a power series expansion in terms
of the strong-force coupling constant $\alpha_s$
(which is smaller than 1);
the difference between LO, NLO and NNLO
calculations is primarily 
that of considering the following level
in this power series expansion,
which leads to a roughly factorial increase
in computational complexity.
It should be pointed out,
nevertheless, 
that NLO calculations for 
simple processes with low 
final state multiplicities
may be computationally cheaper 
than LO calculations for
complex processes with high 
final state multiplicities.
In summary,
it would thus seem that
the intrinsic cost per event 
$\xvec$ of computing $\fxvec$
is to some extent incompressible,
because of the relatively fixed amount 
of arithmetic calculations 
that this involves.
One of the only obvious strategies 
for reducing this cost
consists in improving the
efficiency with which 
these arithmetic operations
are performed~on modern computing systems,
for instance through the use
of parallel programming techniques
such as vectorization or GPU programming,
as discussed later in Sec.~\ref{sec:gpus}.
In addition,
radically new approaches
are also being worked on,
involving for example
the approximation of matrix element
calculations using Machine Learning (ML) 
regression 
\mbox{methods~\cite{bib:bishara,bib:danziger2020}}.

\fussy
\subsection{Inefficiencies in unweighted event generation}
\label{sec:inefficiencies}
The complex workflow described above
presents several challenges 
and opportunities for improvement.
To start with,
there are many sources of inefficiency 
in unweighted event generation,
as discussed in the following.

\sloppy
\paragraph{Phase space sampling inefficiency.} 
The algorithm used for phase space
sampling is the most critical ingredient 
for efficient unweighted event generation.
Some basic techniques,
such as stratified sampling,
which essentially consists
in binning the phase space,
and importance sampling,
which is often implemented
as a change of variables
to parametrize the phase space,
date back to more than 40 years 
ago~\cite{bib:james1980}.
Many algorithms, most notably 
VEGAS~\cite{bib:vegas1978,bib:vegas1980}
or MISER~\cite{bib:miser},
are adaptive, i.e. recursive,
in that their parameters 
are tuned iteratively
as the shape of $\fxvec$ is learnt
by randomly drawing more and more 
phase space points.
Adaptive multi-channel 
algorithms~\cite{bib:multichannel,bib:ohl}
are often used to address
the complex peaking structures
of LHC processes,
by defining the sampling function $\gxvec$
as a weighted sum of functions,
each of which essentially 
describes a different peak.
Many generic sampling algorithms exist,
including very simple ones like 
RAMBO~\cite{bib:rambo},
others derived from VEGAS such as 
BASES/SPRING~\cite{bib:basesspring1,bib:basesspring2}
or MINT~\cite{bib:mint},
and cellular algorithms like
FOAM~\cite{bib:foam}.
Other sampling algorithms
have been developed specifically
for a given generator:
examples include MadEvent~\cite{bib:madevent}
and VAMP~\cite{bib:brass},
which are based on modified versions of VEGAS
and are used in the MG5\_aMC 
and WHIZARD~\cite{bib:kilian}
generators, respectively,
as well as COMIX~\cite{bib:comix},
which is used in Sherpa.

In general, the larger 
the dimensionality of the phase space, 
the lower the unweighting efficiency 
that can be achieved: 
in W+jets at LO, for instance, 
the Sherpa efficiency~\cite{bib:gao2020b}
is 30\% for W+0~jets 
and 0.1\% for~W+4~jets. 
This is an area where research 
is very active, 
and should be actively encouraged, 
as significant cost reductions 
in WLCG compute budgets could be achieved. 
Improvements in this area can only start 
from physics-motivated approaches based 
on the knowledge of phase space peaks, 
but they can be complemented 
by brute-force ML algorithmic 
\mbox{methods~\cite{bib:bendavid,bib:klimek,bib:mlcwp,bib:bothmann2020,bib:gao2020a,bib:gao2020b}}, 
therefore people with different profiles 
can contribute to this area.
The use of one of these ML tools,
Generative Adversarial Networks (GAN),
is being 
investigated~\cite{bib:ml4jets2020}
not only as a way to provide
a more efficient phase space sampling,
but also as a possible replacement
for unweighted event generation altogether,
for example when complemented with 
maximum mean discrepancy
\mbox{methods~\cite{bib:butter2019}}.

\fussy
In this context,
it is useful to point out
that maximizing the efficiency
of unweighted event generation
and minimizing the variance
on total cross section predictions by MC integration
represent two different,
even if closely related, 
strategies for the optimization
of the phase space sampling algorithm.
The two strategies imply 
the use of different loss metrics
during the learning phase of an algorithm, 
and result in different weight distributions.
This is discussed in detail 
in Ref.~\cite{bib:foam},
and to some extent also 
in Ref.~\cite{bib:brass}.
A completely different optimization 
strategy~\cite{bib:oasis}
for the sampling algorithm 
has also been recently proposed,
where the goal is that 
of giving priority to populating
the regions of phase space
which are most sensitive 
to the presence of a signal
or to the value of a parameter.

There are several reasons for the very 
large set of sampling codes. 
Many of them represent evolutions of VEGAS, 
others are completely different algorithms 
(like FOAM), and many of the modern ones 
are based on ML techniques. 
Some of these codes exist for historical reasons,
because of the different choices adopted 
over time by each time. 
Possible work on a ``common'' 
integrator is sometimes mentioned in the community. 
Another possible way forward would consist 
in trying to harmonise the software interfaces 
of these packages, so that each generator 
could plug in different sampling algorithms 
and implementations. 
Discussions in this direction 
have already started 
in the context of the ongoing developments 
on GPU ports and on ML algorithms. 

\sloppy
\newcommand{\kT}{$\mathrm{k}_\mathrm{T}$}
\newcommand{\pT}{$\mathrm{p}_\mathrm{T}$}
\paragraph{Slicing and biasing.}
A further issue~\cite{bib:biasing},
somewhat related to sampling inefficiencies, 
is that jet production cross sections 
fall very sharply as the transverse momenta (\pT) 
of the leading jets increase,
and generating events with uniform weight 
generally fails to give a reasonable
yield in the high-\pT\ regions of phase space. 
One approach to solving this problem (``slicing'') 
is to produce several independent samples of events, 
using different generation cuts in each one, 
in order to populate all the regions of interest. 
An additional approach 
(``biasing'' or ``enhancement''),
available for instance in 
POWHEG~\cite{bib:biasing},
MG5\_aMC~\cite{bib:biasingmg1,bib:biasingmg2},
Sherpa~\cite{bib:sherpaenhanced},
Pythia8~\cite{bib:biasingpythia}
and Herwig7.1~\cite{bib:herwig71},
consists in generating samples 
of events with non uniform weights,
the shape of whose distribution
can however be controlled by 
user-defined suppression factors.
Both approaches are used in practice
by the LHC experiments,
as each has its pros and cons,
and both reduce the resources required 
to populate the low-statistics 
tails of distributions. 
With additional work, 
these methods could help reduce 
the overall event generation 
resource requirements \mbox{at HL-LHC}.

\fussy
\paragraph{Merging inefficiency.} 
Merging prescriptions
(e.g. MLM, CKKW-L at LO,
and FxFx, MEPS@NLO at NLO) 
imply the rejection of some events 
to avoid double\break counting,
between events produced 
with $n$+1 jets 
in the matrix element,
and events produced
with $n$ jets 
in the matrix element 
and one jet from the parton 
shower~\cite{bib:alwall08}.
This is only needed 
if the required final state
includes a variable number 
of jets
$\njets$ between 0 and $n$,
i.e. for so-called ``merged''
or ``multi-leg'' setups.
The resulting inefficiencies
can be relatively low depending on the process, 
but they are unavoidable 
in the algorithmic strategy 
used by the underlying 
physics modeling.
The merging efficiency 
of the MLM prescription, for instance, 
is discussed in in Ref.~\cite{bib:alwall09},
which~shows~how this can be improved
using a method like \mbox{shower-\kT\ MLM}.

\fussy
\paragraph{Filtering inefficiency.} 
An additional large source 
of inefficiency is due 
to the way the experiments 
simulate some processes, 
where they generate large 
inclusive event samples, 
which are then filtered 
on final-state criteria 
to decide which events 
are passed on to detector simulation 
and reconstruction 
(e.g. CMS simulations 
of specific $\mathrm{\Lambda}_\mathrm{B}$ decays 
have a 0.01\% efficiency,
and ATLAS B-hadron filtering 
in a V+jets sample has $\sim$10\%
efficiency~\cite{bib:ATL-PHYS-PUB-2017-006}). 
This inefficiency could be reduced 
by developing filtering tools 
within the generators themselves, 
designed for compatibility 
with the requirements of the experiments. 
A particularly wasteful example 
is where events are separated 
into orthogonal subsamples by filtering, 
in which case the same large inclusive sample 
is generated many times, 
once for each filtering stream: 
allowing a single inclusive event generation 
to be filtered into several 
orthogonal output streams 
would improve~efficiency.
Filtering is an area where 
the LHCb collaboration
has a lot of experience~\cite{bib:ilten} 
and has already obtained 
significant speedups through various techniques. 
In this context, one should also note that
the speed of color reconnection 
algorithms~\cite{bib:cr1,bib:cr2}
is a limiting factor for simulating 
rare hadron decays in~LHCb.

\fussy
\paragraph{Sample sharing.}
In addition to removing inefficiencies, 
other ways could be explored 
to make maximal use of the CPU spent 
for generation by reusing samples 
for more than one purpose. 
Sharing parton-level, or even particle-level, samples 
between ATLAS and CMS 
is being discussed 
for some physics analyses. 
However, the implications of the 
statistical correlations that this would introduce
need further investigation 
in the context of combinations 
of results across experiments.

\sloppy
\paragraph{Sample reweighting.}
Another way to re-use samples 
is through event reweighting. 
Recently,
there have been 
major improvements
in available tools in this
area~\cite{bib:gainer2014,bib:mrenna,bib:bothmann16,bib:mg5amc,bib:mattelaer2016,bib:LH17},
which have made it possible
to obtain systematic uncertainty variations 
as well as
reweighting to alternative model parameters.
The latter may be useful for example 
in new physics searches, 
but also in the optimization
of experimental measurements 
of model parameters~\cite{bib:valassiwdr}.
This machinery is particularly important 
because in the past 
obtaining these variations 
would have required 
multiple samples to go through 
detector simulation and reconstruction, 
whereas the reweighting only requires 
this overhead for a single sample 
that can then be reused in multiple ways. 
This significantly reduces 
the CPU and storage requirements 
for the same end result. 
However, this issue can still 
be explored further 
as in some areas there are limitations 
to the validity of these reweighting 
schemes~\cite{bib:bothmann16,bib:mattelaer2016,bib:LH17,bib:hwreweight1,bib:hwreweight2}.
In addition, 
some systematic uncertainty variations, 
such as merging scale variations, 
are not yet available as weights 
but there is work ongoing. 
There are also systematic variations 
such as changes of the hadronization model 
which are not well suited to the type 
of event reweighting discussed here, 
but for which alternative approaches 
using ML techniques 
to train an ad-hoc reweighting between samples
are under 
\mbox{investigation~\cite{bib:atlasreweight,bib:cranmer2015,bib:andreassen,bib:peyre,bib:rogoz}}.

\newcommand{\bbbar}{b$\bar{\mathrm{b}}$}
\newcommand{\ttbar}{t$\bar{\mathrm{t}}$}
\paragraph{Negative weights.} 
In NLO calculations,
matching prescriptions 
(e.g. MC@NLO, POWHEG, etc.) 
are required 
to avoid double counting 
between phase space configurations
that may come both from $\mathbb{H}$-events 
and from $\mathbb{S}$-events with parton showers.
The solution of this issue becomes 
technically even more complex at the NNLO.
A widely used NLO matching prescription, 
MC@NLO~\cite{bib:mcnlo}, 
is implemented 
by using a ``modified subtraction method''
that may lead to the appearance 
of events with negative weights. 
A MC unweighting procedure is still applied, 
but the resulting events are ``unweighted'' 
in the sense that their weight 
can only be +1 or -1. 
This is a source of (possibly large) inefficiency, 
as larger event samples must be generated 
and passed through the experiment simulation, 
reconstruction and analysis software, 
increasing the compute and storage requirements.
For a fraction $r$ 
of events with weight~\mbox{-1},
the number of events to generate 
increases by a factor \mbox{1/(1-2$r$)$^2$},
because the statistical error on MC predictions 
is a factor \mbox{1/(1-2$r$)} higher;
for a more detailed explanation of these formulas,
see for instance Ref.~\cite{bib:frederix2020}.
For example, 
negative weight fractions equal to 
$r$=25\% and $r$=40\%, which may be 
regarded as worst-case scenarios 
occurring in \ttbar\ and 
H\bbbar\ production~\cite{bib:frederix2020}, 
respectively, imply the need 
to generate 4 times and 25 times 
as many~events 
to achieve the same statistical 
precision on MC predictions.

Negative weights can instead be 
almost completely
avoided, by design, 
in another popular NLO matching prescription, 
POWHEG~\cite{bib:POWHEG},
which however is only available
for a limited number of processes.
POWHEG describes the relevant physics 
in a different way with respect to MC@NLO, 
so that predictions which have formally 
the same level of accuracy 
may visibly differ in the two codes, 
and are associated with different systematics  
(see Ref.~\cite{bib:frederix2020} 
for an \mbox{in-depth discussion}).
Negative weights can also
be avoided in the
KrkNLO~\cite{bib:krknlo}
matching prescription,
which is based on a very different approach
from those used by MC@NLO and POWHEG;
this method however is only available
for a limited number of processes,
and so far has been rarely used 
in practice by the LHC experiments.

Progress in this area 
at the fundamental physics level
can only be achieved by theorists,
and research is active in this area.
For instance, 
a modified MC@NLO matching procedure
with reduced negative weights,
known as \mbox{MC@NLO-$\mathrm{\Delta}$},
has recently been proposed~\cite{bib:frederix2020}.
Similarly, techniques to significantly 
reduce the negative weight fraction 
are also available in 
Sherpa~\cite{bib:danziger2020}. 
Negative weights also exist 
for NNLO calculations,
for instance in the UN2LOPS~\mbox{prescription~\cite{bib:un2lops}}.

It should be stressed that
negative weights due to matching 
are absent in LO calculations.
One possibility for avoiding negative weights,
while possibly still achieving a precision beyond LO,
could then be to generate LO multi-leg setups
and reweight them to higher order predictions;
a careful evaluation
of the theoretical accuracy of this procedure
would however be needed in this case.
In addition,
negative weights can also happen at LO 
because of not-definite-positive 
parton distribution function sets 
and interference terms, 
which is particularly relevant for
effective field \mbox{theory calculations}. 

One should also note that
developments to incorporate contributions 
in parton shower algorithms beyond
the currently adopted approximations, 
see e.g. Refs.\cite{bib:triple,bib:platzer2018,bib:martinez2018}, 
very often necessitate 
weighted evolution algorithms.
In the future,
this may represent another mechanism
leading to the appearance
of events with negative weights,
in addition to and distinct 
from NLO matching prescriptions.
Overcoming the prohibitively 
broad weight distributions 
is subject to an ongoing development 
and might necessitate structural changes 
in the event generation 
workflow. An example is the resampling 
approach proposed in Ref.~\cite{bib:olsson2019},
which also contains a useful historical
review of sampling/importance 
resampling~\cite{bib:sir} techniques
in the broader context 
of Monte Carlo~\mbox{simulations}.

The term resampling has 
also been used to indicate the unrelated technique
of positive resampling~\cite{bib:posresampler},
which has recently been proposed 
as a new approach~for 
reducing the impact~of negative weights
introduced by NLO matching prescriptions.
The idea behind this method,
which addresses negative weights
as a statistics problem without 
looking at the underlying theoretical physics,
is to remove negative weights using 
a quasi-local weight rebalancing scheme.
While positive resampling
uses histograms to determine 
bin-by-bin reweighting factors,
a neural resampling~\cite{bib:nnresampler} approach
has later been proposed as an extension of this method,
using neural networks to determine reweighting factors
in the unbinned high-dimensional phase space.

\fussy
\subsection{Accounting of compute budgets for generators}
While progress has been made 
in the HSF generator WG
to better understand which areas 
of generator software 
have the highest computational cost, 
more detailed accounting 
of the experiment workloads 
and profiling of the main 
generator software packages 
would help to further 
refine R\&D priorities.

\paragraph{Accounting of CPU budgets in ATLAS/CMS.}
Thanks to a large effort 
from the generator teams 
in both experiments, 
a lot of insight 
into the settings used to support 
each experiment's physics programme 
was gained within the WG.
It is now clear that 
the fraction of CPU 
that ATLAS spends for event generation 
is somewhat higher than that in CMS,
although the difference is lower 
than previously thought:
the latest preliminary estimates
of these numbers are 12\% and 5\%, respectively.
A more detailed study 
of the different strategies is ongoing,
in particular by analysing 
individually the CPU costs 
of the main processes simulated 
(notably, V+jets, \ttbar, 
diboson and multijet). 
This effort aims at providing
these figures also as absolute numbers
in normalized HEP-SPEC06 (HS06) 
seconds~\cite{bib:hs06,bib:bmk2020}, 
to allow a more meaningful comparison
of the compute budgets for event generation
in ATLAS and CMS.

A practical issue is that 
these figures had to be harvested 
from logs and production system databases 
a posteriori.
Deriving precise numbers for CMS 
has been particularly difficult, 
requiring significant person hours 
to extract the required information,
as until recently the event generation (GEN) 
and detector simulation (SIM) steps 
were mixed in a single software application, 
and no separate accounting figures for GEN and SIM 
could be recovered from past job logs,
therefore Grid costs 
had to be extrapolated 
from ad-hoc local tests.
CMS is now also using workflows  
including GEN-only applications, 
like that used in ATLAS,
which makes GEN CPU accounting easier.
In addition, job monitoring information
in CMS is presently kept for only 18 months,
which complicates the analysis of past productions,
and does not always include reliable 
HEP-SPEC06 metrics.
For the future, it would be important to 
establish better mechanisms 
to collect all this information, 
to allow for an easy comparison 
between different experiments. 
It would also help 
if the various types 
of efficiencies described above 
(sampling, merging and filtering) 
could be more easily retrieved 
for all simulated~\mbox{processes}. 

\paragraph{Profiling of generators using production setups.\!}
Another area where the WG has been active, 
but more work is needed, 
is the definition and profiling 
of standard generator setups, 
reproducing those used in production. 
This has been used to compare 
the speeds of Sherpa and MG5\_aMC 
in the configurations used 
by ATLAS and CMS, respectively. 
For instance, Sherpa was found 
to be 3 to 8 times slower than MG5\_aMC 
in the generation of NLO W+(0-2)~jets, 
but the exact ratio 
depends on some of the model parameters 
used in Sherpa, 
e.g. the dynamical scale choice of Sherpa,
which results in taking about 
50\% of the total CPU time for generation: 
when modifying Sherpa 
to use an equivalent scale to MG5\_aMC, 
the CPU consumption for this process 
was reduced by over a factor of two. 
The choice of a scale, however,
has important consequences not only 
on computational costs, 
but also on physics accuracy: 
an in-depth discussion 
of this important issue, 
which has been described in 
many research papers by different 
teams of theorists (see, for instance, Refs.~\cite{bib:ckkw,bib:ckkwl,bib:hoeche2012,bib:hoeche2013,bib:frederix2020}), 
is beyond the scope of this paper, 
but the WG will continue 
to investigate the computing 
and physics implications of such choices.

Detailed profiling 
of different generator setups 
has also already helped 
to assess the CPU cost of external PDF libraries, 
and to optimise their 
use~\cite{bib:konstantinov,bib:martin}.
The profiling of the memory footprint 
of the software would also be very useful,
and may motivate in some cases a move 
to multithreading or multiprocessing approaches.

\subsection{Modernisation of generator software}
\label{sec:gpus}
More generally, 
as is the case for many software packages 
in other areas of HEP, 
some R\&D on generators 
would certainly be needed 
to modernise the software 
and make it more efficient, 
or even port it to more modern 
computing architectures
(see also the discussion of these issues
in the Snowmass 2013 
report~\cite{bib:snowmass2013}
and in the HSF~CWP~\cite{bib:hsfcwp}).

\sloppy
\paragraph{Data parallelism, GPUs and vectorization.} 
The data flow of an MC generator, 
where the same function $\fxvec$,
corresponding to the matrix element
for the simulated HEP process,
has to be computed over and over again 
at many phase space points $\xvec_i$
(i.e. for many different events), 
should, in principle, lend itself naturally 
to the data parallel approaches 
found in GPU compute kernels,
and possibly to some extent 
in CPU vectorized code.
In other words,
event-level parallelism
looks like an appropriate approach
to try and exploit efficiently
these architectures~\cite{bib:lhcctalk}.
In this respect,
generators should
be somewhat easier 
to reengineer efficiently for GPUs
than detector simulation software
(notably Geant4~\cite{bib:geant4}),
where the abundance
of conditional branching
of a stochastic nature
may lead to ``thread divergence''
and poor software performance
(see, for examples,
Refs.~\cite{bib:seiskari,bib:g4cu,bib:geantvreport,bib:canal2019,bib:gheata2019,bib:geantv}).

Porting and optimizing generators on GPUs 
is especially important to be able 
to exploit modern GPU-based HPCs 
(such as SUMMIT~\cite{bib:summit}, 
where 95\% of the compute capacity 
comes from GPUs~\cite{bib:summit95}).
Some work in this direction 
was done in the past on MG5\_aMC, 
including
both a port to GPUs
(HEGET~\cite{bib:heget1,bib:heget2,bib:heget3})
of the library that was used in MG5\_aMC,
before ALOHA~\cite{bib:aloha} was introduced,
for the automatic generation 
of matrix element code
(HELAS~\cite{bib:helas1,bib:helas2}),
and a port to GPUs of VEGAS and BASES
(gVEGAS and gBASES~\cite{bib:kanzaki1,bib:kanzaki2}).
This effort,
which unfortunately never reached production quality,
is now being revamped 
by the WG~\cite{bib:lhcctalk,bib:roiser20}, 
in collaboration with the MG5\_aMC team, 
and represents one of the main 
R\&D priorities of the WG. 
This work is presently focusing on Nvidia CUDA, 
but abstraction libraries
like Alpaka~\cite{bib:alpaka,bib:alpakagit}
or oneAPI~\cite{bib:oneapi}
will also be \mbox{investigated}. 

GPUs may also be relevant 
to the ML-based phase space sampling algorithms
discussed in Section~\ref{sec:inefficiencies};
some recent work in this area
has targeted GPUs 
explicitly~\cite{bib:zmc,bib:vegasflow}.
Similar techniques involving ML techniques on GPUs
have recently been used for the computation 
of parton distribution functions~\cite{bib:pdfflow},
which are another essential building block
of the event generator software chain for LHC processes.
Finally, work is also 
ongoing~\cite{bib:gpurng}
on the efficient exploitation of GPUs 
in the pseudo-random number generation libraries 
that are used in all MC generators
(see Ref.~\cite{bib:rng} for a 
recent review of these~components).

\fussy
\paragraph{Task parallelism, 
multithreading, multiprocessing.} 
The use of concurrency mechanisms 
based on multiprocessing (MP) 
or multithreading (MT) 
within event generators 
has increased in recent years, 
but it is not yet a common practice. 
Most often, the use of single-threaded 
(ST), single-process (SP), executables
is not a problem, as the memory footprint 
of event generation is small and 
usually fits within the 2 GB per core 
available~on WLCG nodes, 
which makes it possible to exploit 
all of the available 
cores by running 
many independent
applications in parallel. 
However, there are cases 
(e.g. diboson production,
or Z and Z$\gamma$+jets 
production with electroweak corrections, 
all with up to 4 additional jets) 
where more than 2 GB of memory,
and even as much as 4 GB,
may be needed;
this leads to \mbox{inefficiencies} 
as some CPU cores remain unused,
which could be avoided using MT or MP approaches 
to reduce memory footprints. 
Very often, 
the experiments do not use generators 
as standalone applications, 
but instead embed them in their own 
event processing frameworks, 
which may themselves implement 
MP or MT approaches. 
For MT frameworks, the fact
that some generator packages 
(such as \mbox{EvtGen}~\cite{bib:evtgenmk})
are not thread safe 
may also lead to inefficiencies, 
as this often implies that access 
to these components must be locked 
and their methods can only be used 
by one thread at a time. 
Many different use cases and approaches 
exist in the various experiments, 
as described more in detail~\mbox{below}.

\fussy
In ATLAS, the most commonly used 
workflow for event generation
currently consists in executing 
several independent ST/SP applications
based on the experiment's 
event processing framework, 
Athena~\cite{bib:athena}, 
each running as an independent Grid job 
on a different CPU core. 
Less frequently,
ATLAS also uses multi-core jobs, 
using either ad-hoc features in Athena or,
in the specific case of MG5\_aMC,
the internal concurrency mechanism
provided by this generator;
neither of these options, however,
leads to an overall reduction
in memory footprint,
as they both
ultimately consist in 
spawning several 
independent 
ST applications
on the available cores.
The ATLAS event processing framework, 
Athena~\cite{bib:athena}, 
does have a MP extension based 
on forking and copy-on-write, 
AthenaMP~\cite{bib:athenamp1,bib:athenamp2},
which is routinely used to reduce 
per-core memory footprints of the ATLAS 
simulation and reconstruction workflows, 
but currently this is not used 
for any GEN workflows.
ATLAS is also making progress
in the development of a fully multi-threaded
event processing framework,
AthenaMT~\cite{bib:athenamt},
but this effort is also mainly 
focusing on simulation~\cite{bib:atlassim}
and reconstruction workflows
rather than on event generation.
In particular, as in the 
similar MT developments in LHCb and CMS,
described below,
one of the main aims of this work
is the integration 
into the experiment's workflows
of the recent multi-threaded version
of the Geant4 detector simulation 
toolkit, Geant4-MT~\cite{bib:geant4mt}.

\sloppy
In LHCb, event generation 
currently proceeds only via ST/SP Grid jobs.
A notable difference with respect 
to the ATLAS GEN-only jobs
is that LHCb uses a GEN-SIM workflow 
where the same application, 
Gauss~\cite{bib:gauss},
based on the Gaudi~\cite{bib:gaudi1,bib:gaudi2} 
event processing framework,
performs both the event generation 
and detector simulation steps. 
To improve the efficiency 
of these workflows, 
LHCb is gradually moving away from ST/SP Gauss. 
An MP framework using 
forking and copy-on-write
as in AthenaMP, GaussMP,
was recently used for some MC productions, 
but this was only a temporary ad-hoc solution
for the low-memory many-core 
Intel Knights Landing (KNL) CPUs 
deployed on the Marconi HPC system 
at CINECA~\cite{bib:gaussmp}.
In the future, LHCb plans to replace 
its current SP/ST application
by a fully multi-threaded version of Gauss
based on an experiment-independent
GEN-SIM framework, Gaussino~\cite{bib:gaussino},
which is built directly on Gaudi and is
interfaced to Geant4-MT.
Gaussino, whose development
is making rapid progress,
has the potential for a much better 
optimization of memory usage, 
especially in the SIM step.
To achieve thread safety,
Gaussino uses a single-threaded
locking instance of EvtGen
to handle decays;
special care is also taken 
in the way Gaussino is interfaced to Pythia8, 
as described 
in Ref.~\cite{bib:gaussino}.

In CMS, event generation
is embedded within the multi-threaded 
C++ event processing framework, 
CMSSW~\cite{bib:cmssw0,bib:cmssw1,bib:cmssw2}. 
All CMS workflows for event generation 
involve multi-core Grid jobs
(either GEN-SIM or, more recently, GEN-only), 
where a single instance of the CMSSW application 
simultaneously uses several CPU cores.
One of the primary motivations for CMS 
to use a MT framework is to reduce the amount
of memory used per CPU core~\cite{bib:cmssw1},
a goal that is particularly important, 
and has been achieved in production
(also thanks to the use of the new GEANT4-MT),
for GEN-SIM workflows~\cite{bib:cmsmem}.
Some examples of the integration of 
event generators 
in the CMS worflow, and more particularly
in the MT CMSSW framework,
are described in Refs.~\cite{bib:bendavidcms,bib:cmstwiki}.
The simplest use case is that where
a general-purpose generator
(like Pythia8, Herwig7, or Sherpa)
is used both for the generation 
of parton-level events
and for their hadronization and decay.
In this case, the CMSSW main thread 
starts many separate worker processes 
on the available CPU cores,
ensuring that each worker receives
the appropriate random numbers
to process the event assigned to it
(a mechanism which has some similarities
to that used in GaussMP by LHCb).
A second important use case is that where
event generation is split into two steps,
the generation of parton-level events
in the Les Houches Event File (LHEF) 
format~\cite{bib:lha01,bib:lhef}
using a matrix-element generator
(like MG5\_aMC or POWHEG),
and the hadronization 
and decay to stable particles
of those parton-level events
using other tools 
(like Pythia8, Herwig7, EvtGen or Tauola).
To execute the LHEF event generation step
concurrently on the available cores, 
CMSSW provides a generic mechanism
where several 
independent 
processes are spawned on the available cores;
in the case of MG5\_aMC,
its internal concurrency mode 
may also be used, but this also consists
in spawning several ST applications,
as already mentioned. 
The concurrent execution of the hadronization 
and decay step, conversely, 
is always handled by CMSSW 
using its internal multi-threading: 
the main challenge in this context 
is that 
decayers like EvtGen and Tauola
are not thread-safe
and may only be used to process 
one event at a time~\cite{bib:cmstwiki}.

\sloppy
Multiprocessing approaches
involving several nodes
may also be useful 
to speed up the integration and optimization step 
for complex high-dimensional final states.
In particular,
Sherpa workflows based on the 
Message Passing Interface
(MPI)~\cite{bib:mpi},
which have been available 
for quite a long time,
have been found very useful
by ATLAS and CMS to speed up 
the preparation of integration grids
on local batch clusters.
A lot of work has also been done 
in recent years to implement
and benchmark MPI-based workflows 
on HPC systems.
For instance,
the Sherpa LO-based generation 
of merged many-jet samples 
has been successfully 
tested~\cite{bib:hoeche2019}
on the Cori~\cite{bib:cori}
system at NERSC,
both on traditional Intel Haswell CPUs
and on many-core
Intel KNL CPUs.
This work has used
a technique similar
to that previously developed~\cite{bib:childers2017}
for testing and benchmarking
the scaling of the parallel execution
of Alpgen on Mira~\cite{bib:mira}
at ALCF,
a supercomputer based on IBM PowerPC CPUs.
New event formats, 
migrating LHEF to HDF5, 
have also been instrumental 
in enabling the execution of the 
Sherpa tests at Cori.
MPI integration has also been 
completed for 
MG5\_aMC~\cite{bib:mattelaer2019}.
In this context, 
one should note that,
although HPCs
offer very high-speed
inter-node connectivity, 
HPC resources
can be exploited efficiently 
even without using this connectivity:
in particular,
WLCG workflows,
including generators,
generally use these systems 
as clusters of unrelated nodes,
because the computational workflow
can be split up into independent
tasks on those~nodes.

\sloppy
Hybrid parallelization approaches
are also possible,
where multithreading
or multiprocessing techniques
are used internally 
on a single multi-core node,
while the MPI protocol
is used to manage the 
communication between
distinct computing nodes.
This approach is implemented
for example in the WHIZARD~\cite{bib:brass}
and MCFM~\cite{bib:mcfm} codes,
both of which combine
OpenMP~\cite{bib:omp} multithreading
on individual multi-core nodes 
with MPI message passing between them.

\fussy
\paragraph{Generic code optimizations.}
A speedup of generators 
may also be achievable 
by more generic optimizations, 
not involving concurrency. 
For instance, one could study 
if different compilers 
and build strategies~\cite{bib:mattelaer2019} 
may~lead to any
improvements. 
Another possible strategy 
is to search for redundant calculations,
i.e. to investigate if
some numerical results can be reused 
more than once, for instance via data caching.
Recent studies~\cite{bib:konstantinov}
on~the way LHAPDF6 is used in Pythia8
have indeed resulted in significant speedups
through better data~caching,
and similar studies are in progress
for Sherpa~\cite{bib:martin}.
Ongoing studies~\cite{bib:kiran}
on MG5\_aMC have similarly shown
that important speedups may
obtained through ``helicity recycling'',
i.e. by avoiding the recomputation 
of some building blocks of Feynman diagrams
which are needed in more than
one matrix element calculation.

\section{Physics challenges (increasing precision)}
\label{sec:phys}

\sloppy
In addition to software issues, 
important physics questions 
should also be addressed about 
more accurate theoretical predictions,
above all NNLO 
QCD calculations,
but also electroweak (EW) corrections, 
and their potential impact 
on the computational cost 
of event generators at HL-LHC. 
For a recent review of these issues,
see for example Ref.~\cite{bib:maltoni}. 
Some specific NNLO calculations 
are already available and used today 
by the LHC experiments in their data analysis. 
For example, the measurements 
of fiducial \ttbar\ cross sections,
extrapolated to the full phase space,
are compared to the predictions 
of TOP++~\cite{bib:toppp}, 
accurate to NNLO: 
this program, however, 
does not use MC methods and 
cannot be used 
to generate unweighted events.
Research on NNLO matching
has also made significant progress,
for example on the 
NNLOPS~\cite{bib:minlo_nnlops1}, 
GENEVA~\cite{bib:geneva2}, 
UN2LOPS~\cite{bib:un2lops} and 
MINNLOPS~\cite{bib:minnlops,bib:mazzitelli} 
prescriptions.
In addition, samples of unweighted events 
are routinely generated 
for Higgs boson final states 
using the POWHEG/MINLO 
NNLOPS approach~\cite{bib:minlo_nnlops1,bib:minlo_nnlops2}.
With a view to HL-LHC times, however, 
some open questions remain to be answered, 
as discussed below.

\sloppy
\paragraph{NNLO: status of theoretical physics research.} 
The first question is for which processes 
QCD NNLO precision will be available 
at the time of the HL-LHC. 
For example, 
first results for triphoton results at NNLO
have recently been published~\cite{bib:chawdry}:
when would NNLO 
be expected for other 2$\rightarrow$3 processes 
or even higher multiplicity final states? 
Also, for 
final states such as \ttbar,
differential NNLO predictions 
exist~\cite{bib:czakon2016,bib:grazzini2019}
and the first matched computation for NNLO+PS
was very recently achieved~\cite{bib:mazzitelli},
but the software for generating 
unweighted NNLO+PS events using the latter 
is not yet publicly available
for use in the experiments,
when can this be expected?
In particular, it would be 
important to clarify 
which are the theoretical 
and more practical challenges 
in these calculations, 
and the corresponding computational strategies 
and predicted impact on CPU time needs 
(e.g. more complex definition 
of matching procedures,
higher fraction of negative weights,
and more complex \mbox{2-loop~MEs}?).

The accuracy of shower generators
is also important in this context.
Current shower generators rely 
on first order splitting kernels, 
together with an appropriate scheme 
to handle soft emissions. 
Recent work aims at improving parton showers 
by increasing their accuracy either 
by developing novel shower schemes 
within the standard parton 
or dipole branching, such as 
DIRE~\cite{bib:dire} 
and Vincia~\cite{bib:vincia} 
or by going beyond the typical probabilistic 
approach~\cite{bib:deductor} 
and by incorporating higher order splitting 
functions~\cite{bib:krknlo,bib:triple,bib:nlodglap,bib:dulat}.
In addition, very recently,  
significant theoretical advance opening 
the way to showers with 
next-to-leading logarithmic (NLL) accuracy has been 
achieved~\cite{bib:dasgupta}.

To match NNLO accuracy in QCD, 
EW corrections must also be included. 
Recently, much progress 
has been achieved on the automation 
of the computation of EW 
corrections~\cite{bib:actis2013,bib:kallweit2015,bib:schonherr2018,bib:frederix2018},
to the point that fixed-order 
NLO QCD and EW corrections 
are readily available for any process 
of interest at the LHC. 
A general interface of these calculations 
to shower generators that correctly account 
for QED radiation for these computations, 
however, is not yet available. 

\sloppy
An additional concern,
in general but especially 
in higher-order phenomenology, 
is the control of numerical 
and methodological errors 
at the sub-percent level. 
This is relevant for processes 
where high-precision measurements 
and predictions are available, 
but also to efficiently and 
precisely test the input parameter 
dependence (PDFs, $\alpha_s$, etc.).
These issues, and the way 
in which they are addressed 
in the MCFM parton-level code, 
are discussed in detail 
in Ref.~\cite{bib:mcfm}. 
A key component of this code 
is a fully parallelized 
phase space integration,
using both OpenMP and MPI 
on multi-core machines and cluster setups,
where technical cutoffs can be
controlled at the required 
level \mbox{of precision}.

\fussy
\paragraph{NNLO: experimental requirements at HL-LHC.}
The second question 
is for which final states 
unweighted event generation 
with NNLO precision 
would actually be required 
(\ttbar\ production is a clear candidate), 
and how many events would be needed. 
One should also ask 
if reweighting LO event samples to NNLO 
would not be an acceptable 
cheaper alternative to address 
the experimental needs,
and what would be the theoretical
accuracy reached by this procedure.

\paragraph{Size of unweighted event samples 
required at HL-LHC.}
Another question to be asked, 
unrelated to NNLO, 
is in which regions of phase space 
the number of unweighted events must be
strictly proportional to the luminosity.
For example, in the bulk (low \pT) 
regions of W boson production 
it is probably impossible 
to keep up with the data, 
due to the huge cross section. 
Alternative techniques 
could be investigated, 
to avoid the generation 
of huge samples of unweighted events.

\section{Conclusions}
\label{sec:concl}

This paper has been prepared
by the HSF Physics Event Generator Working Group
as an input to the 
LHCC review of HL-LHC computing,
which has started in May 2020.
We have reviewed the main 
software and computing challenges 
for the Monte Carlo physics event generators 
used by the LHC  experiments, 
in view of the HL-LHC physics programme. 

Out of the many issues 
that we have described,
we have identified 
the following five 
as the main priorities
on which the R\&D 
on the computational aspects of generators, 
and in particular the activities of our WG,
should focus:
\begin{enumerate}
\item
Gain a more detailed understanding 
of the current CPU costs 
by accounting and profiling. 
\item
Survey generator codes 
to understand the best way 
to move to GPUs and vectorized code, 
and prototype the port of the software 
to GPUs using data-parallel paradigms.
\item
Support efforts to optimize 
phase space sampling 
and integration algorithms,
including the use of Machine Learning 
techniques such as neural networks.
\item
Promote research 
on how to reduce the cost associated 
with negative weight events, 
using new theoretical 
or experimental approaches.
\item
Promote collaboration, 
training, funding and career opportunities 
in the generator area.
\end{enumerate}
Additional material
about these and the other issues
described in this paper,
including detailed plots and diagrams,
may be found in the recent
presentation by the HSF Generator WG
to the LHCC~\cite{bib:lhcctalk}.
We~plan to report 
about the progress 
in these areas
during the more in-depth
LHCC review of HL-LHC software,
which is currently scheduled 
in~Q3~2021,
and reassess the WG priorities
for future activities 
at that point in time.

\section*{Acknowledgements}

\sloppy
This work received funding 
from the European Union's Horizon 2020 
research and innovation programme 
as part of the Marie Sk\l odowska-Curie 
Innovative Training Network MCnetITN3 
(grant agreement no. 722104).
This research used resources of the 
Fermi National Accelerator Laboratory 
(Fermilab), a U.S. Department of Energy, 
Office of Science, HEP User Facility.
Fermilab is managed by 
Fermi Research Alliance, LLC (FRA), 
acting under Contract No. DE--AC02--07CH11359.
This work was supported by the 
Laboratory Directed Research 
and Development Program of 
Lawrence Berkeley National Laboratory 
under U.S. Department of Energy 
Contract No. DE-AC02-05CH11231.
The work at Argonne National Laboratory 
was supported by the U.S. Department of Energy, 
Office of Science, 
High Energy Physics Center 
for Computational Excellence (HEP-CCE) program 
under Award Number 0000249313.
F.~Krauss acknowledges funding 
as Royal Society Wolfson Research fellow.
M.~Sch\"onherr is funded 
by the Royal Society 
through a University Research Fellowship.
E.~Yazgan acknowledges funding from National 
Taiwan University grant NTU~109L104019.
A.~Si\'odmok acknowledges support from the 
National Science Centre, Poland 
Grant No. 2019/34/E/ST2/00457.

\fussy


\begin{thebibliography}{}

\bibitem{bib:atlas}
ATLAS Collaboration,
{\em The ATLAS experiment 
at the CERN Large Hadron Collider}, 
J. Instrum. \textbf{3} (2008) S08003.
\doi{10.1088/1748-0221/3/08/S08003}

\bibitem{bib:cms}
CMS Collaboration, 
{\em The CMS experiment at the CERN LHC}, 
J. Instrum. \textbf{3} (2008) S08004.
\doi{10.1088/1748-0221/3/08/S08004}

\bibitem{bib:hllhc}
High-Luminosity Large Hadron Collider (HL-LHC).
\url{https://home.cern/science/accelerators/high-luminosity-lhc}

\bibitem{bib:hsfgen2019}
HSF Physics Event Generator Working Group.
\url{https://hepsoftwarefoundation.org/workinggroups/generators.html}

\bibitem{bib:lhcc}
LHC Experiments Committee.
\url{https://committees.web.cern.ch/lhcc}

\bibitem{bib:lhccreport1}
A. Boehnlein et al.,
{\em HL-LHC Software and Computing Review Panel, 1st Report},
CERN-LHCC-2020-012.
\url{https://cds.cern.ch/record/2725487}

\bibitem{bib:lhcc136}
{\em Minutes of the 136th Meeting of the LHCC},
CERN-LHCC-2018-033. 
\url{https://cds.cern.ch/record/2649242}

\bibitem{bib:lhcc139}
{\em Minutes of the 139th Meeting of the LHCC},
CERN-LHCC-2019-010. 
\url{https://cds.cern.ch/record/2689443}

\bibitem{bib:lhcc140}
{\em Minutes of the 140th Meeting of the LHCC},
CERN-LHCC-2019-016. 
\url{https://cds.cern.ch/record/2702745}

\bibitem{bib:lhcc141}
{\em Minutes of the 141st Meeting of the LHCC},
CERN-LHCC-2020-003. 
\url{https://cds.cern.ch/record/2711432}

\bibitem{bib:alice}
ALICE Collaboration,
{\em The ALICE experiment at the CERN LHC}, 
J. Instrum. \textbf{3} (2008) S08002.
\doi{10.1088/1748-0221/3/08/S08002}

\bibitem{bib:lhcb}
LHCb Collaboration, 
{\em The LHCb detector at the LHC},
J. Instrum. \textbf{3} (2008) S08005.
\doi{10.1088/1748-0221/3/08/S08005}

\bibitem{bib:hsflhcc}
HEP Software Foundation,
{\em Common Tools and Community Software}, 
input for the LHCC review of HL-LHC computing, 
HSF-DOC-2020-1 (2020).
\doi{10.5281/zenodo.3779249}

\bibitem{bib:hsfcwp}
HEP Software Foundation,
{\em A Roadmap for HEP Software and Computing R\&D for the 2020s},
Comput. Softw. Big Sci. \textbf{3} (2019) 7.
\doi{10.1007/s41781-018-0018-8}

\bibitem{bib:buckley2019mcnet}
A. Buckley et al.,
{\em Monte Carlo event generators 
for high energy particle physics event simulation},
MCnet-19-02 (2019).
\arxiv{1902.01674}

\bibitem{bib:espp}
CERN Council Open Symposium 
on the Update of European 
Strategy for Particle Physics,
Granada (2019).
\url{https://cafpe.ugr.es/eppsu2019}

\bibitem{bib:wlcg}
Worldwide LHC Computing Grid.
\url{https://wlcg.web.cern.ch}

\bibitem{bib:hatlas}
ATLAS Collaboration,
{\em Combined measurements 
of Higgs boson production and decay 
using up to 80 fb$^{-1}$ 
of proton-proton collision data 
at $\sqrt{s}$=13 TeV collected 
with the ATLAS experiment},
Phys. Rev. D \textbf{101} (2020) 012002.
\doi{10.1103/PhysRevD.101.012002}

\bibitem{bib:hcms}
CMS Collaboration,
{\em Observation of Higgs boson 
decay to bottom quarks},
Phys. Rev. Lett. \textbf{121} (2018) 121801.
\doi{10.1103/PhysRevLett.121.121801}

\bibitem{bib:yellow1}
P. Azzi (ed.) et al.,
{\em Report from Working Group 1:
Standard Model Physics 
at the HL-LHC and HE-LHC},
CERN-LPCC-2018-03,
HL/HE-LHC Workshop, CERN (2018).
\arxiv{1902.04070}

\bibitem{bib:yellow2}
M. Cepeda (ed.) et al.,
{\em Report from Working Group 2:
Higgs physics 
at the HL-LHC and HE-LHC},
CERN-LPCC-2018-04,
HL/HE-LHC Workshop, CERN (2018).
\arxiv{1902.00134}

\bibitem{bib:yellow3}
X. Cid Vidal (ed.) et al.,
{\em Report from Working Group 3:
Beyond the Standard Model Physics 
at the HL-LHC and HE-LHC},
CERN-LPCC-2018-05,
HL/HE-LHC Workshop, CERN (2018).
\arxiv{1812.07831}

\bibitem{bib:yellow4}
A. Cerri (ed.) et al.,
{\em Report from Working Group 4:
Opportunities in Flavour Physics 
at the HL-LHC and HE-LHC},
CERN-LPCC-2018-06,
HL/HE-LHC Workshop, CERN (2018).
\arxiv{1812.07638}

\bibitem{bib:hsfgen2017}
R. Boughezal et al.,
{\em Generator and Theory Working Group Chapter for CWP},
unpublished draft (2017).
\url{https://github.com/HSF/documents/tree/master/CWP/papers/HSF-CWP-2017-11\_generators}

\bibitem{bib:maltoni}
F. Maltoni, M. Sch\"onherr, P. Nason,
{\em Monte Carlo generators}, 
in Ref.~\cite{bib:yellow1}.

\bibitem{bib:hsfgen2018}
HSF Physics Event Generator Computing Workshop, 
CERN (2018).
\url{https://indico.cern.ch/event/751693}

\bibitem{bib:buckley2019acat}
A. Buckley,
{\em Computational challenges for MC event generation},
Proc. ACAT 2019, Saas Fee, 
J. Phys. Conf. Ser. \textbf{1525} (2030) 012023.
\doi{10.1088/1742-6596/1525/1/012023}

\bibitem{bib:sexton2018}
L. Sexton-Kennedy, G. Stewart, 
{\em CWP challenges and workshop aims}, 
in Ref.~\cite{bib:hsfgen2018}.
\url{https://indico.cern.ch/event/751693/contributions/3182927}



\bibitem{bib:sherpa}
E. Bothmann et al., 
{\em Event generation with Sherpa 2.2}, 
SciPost Phys. \textbf{7} (2019) 034.
\doi{10.21468/SciPostPhys.7.3.034}

\bibitem{bib:mg5amc} 
J. Alwall et al.,
{\em The automated computation of tree-level 
and next-to-leading order differential cross sections, 
and their matching to parton shower simulations}, 
JHEP07(2014)079.
\doi{10.1007/JHEP07(2014)079}

\bibitem{bib:josh2018}
J. McFayden,
{\em ATLAS needs and concerns}, 
in Ref.~\cite{bib:hsfgen2018}.
\url{https://indico.cern.ch/event/751693/contributions/3182932}

\bibitem{bib:efe2018}
E. Yazgan,
{\em CMS needs and concerns}, 
in Ref.~\cite{bib:hsfgen2018}.
\url{https://indico.cern.ch/event/751693/contributions/3182936}

\bibitem{bib:javier19}
ATLAS and CMS Collaborations,
{\em Comparison of ATLAS and CMS 
nominal $\mathrm{t}\mathrm{\bar t}$ 
Monte Carlo simulation for Run 2}, 
CMS-DP-2019-011, CERN (2019).
\url{https://cds.cern.ch/record/2678959}

\bibitem{bib:avgpus}
A. Valassi,
{\em Overview of the GPU efforts 
for WLCG production workloads},
Pre-GDB on benchmarking, CERN (2019).
\url{https://indico.cern.ch/event/739897/contributions/3559134}

\bibitem{bib:buckley2011}
A. Buckley et al.,
{\em General-purpose event generators for LHC physics},
Physics Reports \textbf{504} (2011) 145.
\doi{10.1016/j.physrep.2011.03.005}

\bibitem{bib:sjostrand2012}
T. Sj\"ostrand,
{\em Introduction to Monte Carlo techniques 
in High Energy Physics},
CERN Summer Student Lectures (2012).
\url{https://indico.cern.ch/event/190076}
  
\bibitem{bib:sjostrand2016}
T. Sj\"ostrand,
{\em Status and developments of event generators},
Fourth LHC Physics Conference (LHCP2016), 
Lund (2016).
\arxiv{1608.06425}

\bibitem{bib:POWHEGbox}
S. Alioli et al.,
{\em A general framework for implementing NLO
calculations in shower Monte Carlo programs: 
the POWHEG BOX},
JHEP06(2010)043.
\doi{10.1007/JHEP06(2010)043}

\bibitem{bib:pythia82}
T. Sj\"ostrand et al.,
{\em An introduction to PYTHIA 8.2},
Comp. Phys. Comm. \textbf{191} (2015) 159.
\doi{10.1016/j.cpc.2015.01.024}


\bibitem{bib:herwigpp}
M. B\"ahr et al.,
{\em Herwig++ physics and manual},
Eur. Phys. J. C \textbf{58} (2008) 639.
\doi{10.1140/epjc/s10052-008-0798-9}

\bibitem{bib:herwig70}
J. Bellm et al.,
{\em HERWIG 7.0/HERWIG++ 3.0 release note},
Eur. Phys. J. C \textbf{76} (2016) 196.
\doi{10.1140/epjc/s10052-016-4018-8}

\bibitem{bib:herwig71}
J. Bellm et al.,
{\em Herwig 7.1 Release Note},
CERN-PH-TH-2017-109 (2017). 
\arxiv{1705.06919}

\bibitem{bib:alpgen}
M. L. Mangano et al., 
{\em Alpgen, a generator for hard
multiparton processes in hadronic collisions},
JHEP07(2003)001.
\doi{10.1088/1126-6708/2003/07/001}

\bibitem{bib:ariadne}
L. Lönnblad,
{\em Ariadne version 4: A program for simulation 
of QCD cascades implementing the colour dipole model},
Comp. Phys. Comm. \textbf{71} (1992) 15.
\doi{10.1016/0010-4655(92)90068-A}

\bibitem{bib:atlastune}
ATLAS Collaboration,
{\em Summary of ATLAS Pythia 8 tunes},
ATL-PHYS-PUB-2012-003 (2012).
\url{https://cds.cern.ch/record/1474107}

\bibitem{bib:cmstune}
CMS Collaboration,
{\em Extraction and validation 
of a new set of CMS pythia8 
tunes from underlying-event measurements},
Eur. Phys. J. C \textbf{80} (2020) 4.
\doi{10.1140/epjc/s10052-019-7499-4}

\bibitem{bib:newtunes}
X. Ju et al.,
{\em A novel workflow 
of generator tunings in HPC 
for LHC new physics searches},
in Ref.~\cite{bib:hsfgen2018}.
\url{https://indico.cern.ch/event/751693/contributions/3183027}

\bibitem{bib:autotune}
J. Bellm, L. Gellersen, 
{\em High dimensional parameter 
tuning for event generators},
Eur. Phys. J. C \textbf{80} (2020) 54. 
\doi{10.1140/epjc/s10052-019-7579-5}

\bibitem{bib:fxfx}
R. Frederix, S. Frixione, 
{\em Merging meets matching in MC@NLO}
JHEP12(2012)061. 
\doi{10.1007/JHEP12(2012)061}

\bibitem{bib:mcnlo}
S. Frixione, B. R. Webber,
{\em Matching NLO QCD computations 
and parton shower simulations},
JHEP06(2002)029.
\doi{10.1088/1126-6708/2002/06/029}

\bibitem{bib:POWHEG}
S. Frixione, P. Nason, C. Oleari, 
{\em Matching NLO QCD computations 
with parton shower simulations: 
the POWHEG method}, 
JHEP11(2007)070.
\doi{10.1088/1126-6708/2007/11/070}

\bibitem{bib:krknlo}
S. Jadach et al.,
{\em Matching NLO QCD with parton shower 
in Monte Carlo scheme — the KrkNLO method},
JHEP10(2015)052. 
\doi{10.1007/JHEP10(2015)052}

\bibitem{bib:ckkw}
S. Catani, F. Krauss, R. Kuhn, B. R. Webber,
{\em QCD matrix elements + parton showers},
JHEP11(2001)063.
\doi{10.1088/1126-6708/2001/11/063}

\bibitem{bib:ckkwl}
L. L\"onnblad,
{\em Correcting the colour-dipole 
cascade model with fixed order matrix elements},
JHEP05(2002)046.
\doi{10.1088/1126-6708/2002/05/046}

\bibitem{bib:mlm}
M. L. Mangano, M. Moretti, R. Pittau,
{\em Multijet matrix elements 
and shower evolution in hadronic collisions:
$Wb\bar{b}$~+~n-jets as a case study},
Nucl. Phys. B \textbf{632} (2002) 343.
\doi{10.1016/S0550-3213(02)00249-3}

\bibitem{bib:alwall08}
J. Alwall et al., 
{\em Comparative study of various algorithms 
for the merging of parton showers 
and matrix elements in hadronic collisions},
Eur. Phys. J. C \textbf{53} (2008) 473. 
\doi{10.1140/epjc/s10052-007-0490-5}

\bibitem{bib:mepsnlo}
S. H\"oche, F. Krauss, M. Sch\"onherr,
{\em Uncertainties~in 
MEPS@NLO\,calculations\,of\,h+jets},
Phys.\,Rev.\,D\,\textbf{90}\,(2014) 014012.
\doi{10.1103/PhysRevD.90.014012}

\bibitem{bib:minlo}
K. Hamilton, P. Nason, G. Zanderighi, 
{\em MINLO: multi-scale improved NLO}. 
JHEP10(2012)155. 
\doi{10.1007/JHEP10(2012)155}


\bibitem{bib:unlops}
L. L\"onnblad, S. Prestel, 
{\em Merging multi-leg NLO matrix elements
with parton showers},
JHEP03(2013)166.
\doi{10.1007/JHEP03(2013)166}

\bibitem{bib:platzer2012}
S. Pl\"atzer, S. Gieseke, 
{\em Dipole showers and automated NLO matching in Herwig++},
Eur. Phys. J. C \textbf{72} (2012) 2187. 
\doi{10.1140/epjc/s10052-012-2187-7}

\bibitem{bib:platzer2013}
S. Pl\"atzer, 
{\em Controlling inclusive cross sections 
in parton shower + matrix element merging},
JHEP08(2013)114.
\doi{10.1007/JHEP08(2013)114}

\bibitem{bib:bellm2018}
J. Bellm, S. Gieseke, S. Pl\"atzer,
{\em Merging NLO multi-jet calculations 
with improved unitarization},
Eur. Phys. J. C \textbf{78} (2018) 244. 
\doi{10.1140/epjc/s10052-018-5723-2}

\bibitem{bib:evtgen}
D. J. Lange,
{\em The EvtGen particle decay simulation package},
Nucl. Instrum. Meth. A \textbf{462} (2001) 152.
\doi{10.1016/S0168-9002(01)00089-4}

\bibitem{bib:tauola}
S.\,Jadach, Z.\,Was, R.\,Decker, J.\,H.\,K\"uhn,
{\em The $\tau$ decay~library\,TAUOLA,\,version\,2.4},
Comp.\,Phys.\,Comm.\,\textbf{76} (1993) 361.
\doi{10.1016/0010-4655(93)90061-G}

\bibitem{bib:photos}
P. Golonka, Z. Was, 
{\em PHOTOS Monte Carlo: a precision tool 
for QED corrections in Z and W decays},
Eur.\,Phys.\,J.\,C \textbf{45} (2006) 97. 
\doi{10.1140/epjc/s2005-02396-4}


\bibitem{bib:pdf4lhc}
M. Botje,
{\em The PDF4LHC Working Group 
Interim Recommendations} (2011).
\arxiv{1101.0538v1}

\bibitem{bib:lhapdf}
A. Buckley et al.,
{\em LHAPDF6: parton density access 
in the LHC precision era},
Eur. Phys. J. C \textbf{75} (2015) 132.
\doi{10.1140/epjc/s10052-015-3318-8}

\bibitem{bib:rivet}
A. Buckley et al.,
{\em Rivet user manual},
Comp. Phys. Comm. \textbf{184} (2013) 2803.
\doi{10.1016/j.cpc.2013.05.021}

\bibitem{bib:professor}
A. Buckley et al.,
{\em Systematic event generator 
tuning for the LHC}, 
Eur. Phys. J. C \textbf{65} (2010) 331. 
\doi{10.1140/epjc/s10052-009-1196-7}

\bibitem{bib:hepmc}
M. Dobbs, J. B. Hansen,
{\em The HepMC C++ Monte Carlo event record 
for High Energy Physics},
Comp. Phys. Comm. \textbf{134} (2001) 41.
\doi{10.1016/S0010-4655(00)00189-2}

\bibitem{bib:hepmc3}
A. Buckley et al.,
{\em The HepMC3 event record library 
for Monte Carlo event generators},
Comp. Phys. Comm. \textbf{260} (2021) 107310.
\doi{10.1016/j.cpc.2020.107310}

\bibitem{bib:fastjet}
M. Cacciari, G. P. Salam, G. Soyez,
{\em FastJet user manual}, 
Eur. Phys. J. C \textbf{72} (2012) 1896.
\doi{10.1140/epjc/s10052-012-1896-2}

\bibitem{bib:roiser}
S. Roiser et al.,
{\em The LHCb Distributed Computing Model 
and Operations during LHC Runs 1, 2 and 3},
Proc. ISGC2015, Taipei.
\doi{10.22323/1.239.0005}

\bibitem{bib:caravaglios}
F. Caravaglios, M. Moretti,
{\em An algorithm to compute 
Born scattering amplitudes 
without Feynman graphs},
Phys. Lett. B \textbf{358} (1995) 332.
\doi{10.1016/0370-2693(95)00971-M}

\bibitem{bib:james1968}
F. James,
{\em Monte Carlo phase space},
CERN Yellow Report CERN-68-15 (1968).
\doi{10.5170/CERN-1968-015}

\bibitem{bib:james1980}
F. James,
{\em Monte Carlo theory and practice},
Rep. Progr. Phys. \textbf{43} (1980) 1145.
\doi{10.1088/0034-4885/43/9/002}

\bibitem{bib:rng}
F. James, L. Moneta, 
{\em Review of High-Quality 
Random Number Generators}, 
Comput. Softw. Big Sci. \textbf{4} (2020) 2.
\doi{10.1007/s41781-019-0034-3}

\bibitem{bib:weinzierl2000}
S. Weinzierl,
{\em Introduction to Monte Carlo methods},
NIKHEF-00-012 (2000).
\arxiv{hep-ph/0006269}

\bibitem{bib:quasi}
F. James, J. Hoogland, R.Kleiss,
{\em Multidimensional sampling for 
simulation and integration: measures, 
discrepancies, and quasi-random numbers},
Comp. Phys. Comm. \textbf{99} (1997) 180.
\doi{10.1016/S0010-4655(96)00108-7}


\bibitem{bib:toppp}
M. Czakon, A. Mitov,
{\em Top++: A program for the calculation 
of the top-pair cross-section 
at hadron colliders},
Comp. Phys. Comm. \textbf{185} (2014) 2930.
\doi{10.1016/j.cpc.2014.06.021}

\bibitem{bib:vegas1978}
G. P. Lepage,
{\em A New Algorithm for Adaptive Multidimensional Integration}, 
J. Comp. Phys. \textbf{27} (1978) 192. 
\url{https://doi.org/10.1016/0021-9991(78)90004-9}

\bibitem{bib:vegas1980}
G. P. Lepage,
{\em VEGAS: an adaptive 
multi-dimensional integration program},
Cornell report CLNS-447 (1980).
\url{https://cds.cern.ch/record/123074}


\bibitem{bib:ilten}
P. Ilten,
{\em LHCb needs and concerns}, 
in Ref.~\cite{bib:hsfgen2018}.
\url{https://indico.cern.ch/event/751693/contributions/3182938}

\bibitem{bib:davis}
A. Davis,
{\em Fast Simulation in LHCb}, 
Workshop in Efficient Computing 
for High Energy Physics (ECHEP),
Edinburgh (2020).
\url{https://indico.ph.ed.ac.uk/event/66/contributions/844}

\bibitem{bib:collins89}
J. Collins, D. Soper, G. Sterman,
{\em Factorization of Hard Processes in QCD}, 
Adv. Ser. Direct. High Energy Phys. 
\textbf{5} (1989) 1.
\arxiv{hep-ph/0409313v1}

\bibitem{bib:talk-weinzierl-2012}
S. Weinzierl,
{\em NLO Calculations},
Monte Carlo School, Hamburg (2012).
\url{https://indico.desy.de/indico/event/5064/session/8}

\bibitem{bib:talk-zaro-2015}
M. Zaro,
{\em MadGraph5\_aMC@NLO tutorial},
QCD and event simulation 
for the LHC lectures, Pavia (2015).
\url{https://cp3.irmp.ucl.ac.be/projects/madgraph/wiki/Pavia2015}

\bibitem{bib:talk-luisoni-2019}
G. Luisoni,
{\em An introduction to POWHEG},
Dartmouth-UW Experimental/Theory discussion (2017).
\url{https://indico.cern.ch/event/602457/contributions/2435408}

\bibitem{bib:gridpacks1}
J. Alwall et al.,
{\em New Developments in MadGraph/MadEvent},
Proc. SUSY08, Seoul,
AIP Conf. Proc. \textbf{1078} (2009) 84.
\doi{10.1063/1.3052056}

\bibitem{bib:gridpacks2}
MadGraph, 
Technical details for setting up 
and running the Grid Package.
\url{https://cp3.irmp.ucl.ac.be/projects/madgraph/wiki/GridDevelopment}

\bibitem{bib:sherpagrid}
Sherpa Integration,
Sherpa 2.0.0 Manual (2013).
\url{https://sherpa.hepforge.org/doc/SHERPA-MC-2.0.0.html#Integration}

\bibitem{bib:talk-efe-2018}
E. Yazgan,
{\em Event Generators in CMS},
CMS Heavy Flavor Tagging Workshop, Brussels (2018).
\url{https://indico.cern.ch/event/695320/contributions/2850950}

\bibitem{bib:lange}
D. Lange,
{\em Practical computing considerations}, 
in Ref.~\cite{bib:hsfgen2018}.
\url{https://indico.cern.ch/event/751693/contributions/3182940}

\bibitem{bib:bishara}
F. Bishara, M. Montull,
{\em (Machine) Learning amplitudes 
for faster event generation},
DESY 19-232 (2019).
\arxiv{1912.11055}

\bibitem{bib:danziger2020}
K. Danziger,
{\em Efficiency Improvements 
in Monte Carlo Algorithms 
for High-Multiplicity Processes},
Master-Arbeit Thesis, TU Dresden, 
CERN-THESIS-2020-024 (2020).
\url{https://cds.cern.ch/record/2715727}

\bibitem{bib:miser}
W. H. Press, G. R. Farrar,
{\em Recursive Stratified Sampling 
for Multidimensional Monte Carlo Integration},
Computers in Physics \textbf{4} (1990) 190. 
\doi{10.1063/1.4822899}

\bibitem{bib:multichannel}
R. Kleiss, R. Pittau,
{\em Weight optimization in multichannel Monte Carlo},
Comp. Phys. Comm. \textbf{83} (1994) 141.
\doi{10.1016/0010-4655(94)90043-4}

\bibitem{bib:ohl}
T. Ohl,
{\em Vegas Revisited: Adaptive
Monte Carlo Integration Beyond Factorization},
Comp. Phys. Comm. \textbf{120} 
(1999)\,13.\,\doi{10.1016/S0010-4655(99)00209-X}

\bibitem{bib:rambo}
R. H. Kleiss, W. J. Stirling, S. D. Ellis,
{\em A new Monte Carlo treatment 
of multiparticle phase space at high energies},
Comp. Phys. Comm. \textbf{40} (1986) 359.
\doi{10.1016/0010-4655(86)90119-0}

\bibitem{bib:basesspring1}
S.Kawabata,
{\em A new Monte Carlo event generator 
for high energy physics},
Comp. Phys. Comm. \textbf{41} (1986) 127.
\doi{10.1016/0010-4655(86)90025-1}

\bibitem{bib:basesspring2}
S.Kawabata,
{\em A new version of the multi-dimensional 
integration and event generation 
package BASES/SPRING},
Comp. Phys. Comm. \textbf{88} (1995) 309.
\doi{10.1016/0010-4655(95)00028-E}

\bibitem{bib:mint}
P. Nason,
{\em MINT: a Computer Program 
for Adaptive Monte Carlo Integration 
and Generation of Unweighted Distributions}
Bicocca-FT-07-13 (2007).
\arxiv{0709.2085}

\bibitem{bib:foam}
S. Jadach,
{\em Foam: a general-purpose 
cellular Monte Carlo event generator},
Comp. Phys. Comm. \textbf{152} (2003) 55.
\doi{10.1016/S0010-4655(02)00755-5}

\bibitem{bib:madevent}
F. Maltoni, T. Stelzer,
{\em MadEvent: automatic 
event generation with MadGraph},
JHEP02(2003)027.
\doi{10.1088/1126-6708/2003/02/027}

\bibitem{bib:brass}
S.~Brass, W.~Kilian, J.~Reuter,
{\em Parallel Adaptive Monte Carlo 
Integration with the Event Generator WHIZARD},
Eur. Phys. J. C \textbf{79} (2019) 344.
\doi{10.1140/epjc/s10052-019-6840-2}

\bibitem{bib:kilian}
W.~Kilian, T.~Ohl, J.~Reuter,
{\em WHIZARD: Simulating Multi-Particle 
Processes at LHC and ILC},
Eur. Phys. J. C \textbf{71} (2011) 1742.
\doi{10.1140/epjc/s10052-011-1742-y}

\bibitem{bib:comix}
T. Gleisberg, S. H\"oche,
{\em Comix, a new matrix element generator},
JHEP12(2008)039.
\doi{10.1088/1126-6708/2008/12/039}

\bibitem{bib:gao2020b}
C. Gao et al.,
{\em Event generation with normalizing flows},
Phys. Rev. D \textbf{101} (2020) 076002.
\doi{10.1103/PhysRevD.101.076002}

\bibitem{bib:gao2020a}
C. Gao et al.,
{\em i-flow: High-dimensional Integration 
and Sampling with Normalizing Flows},
Mach. Learn.: Sci. \mbox{Technol.} 
\textbf{1} (2020) 045023.
\doi{10.1088/2632-2153/abab62} 

\bibitem{bib:bendavid}
J. Bendavid, 
{\em Efficient Monte Carlo Integration 
Using Boosted Decision Trees 
and Generative Deep Neural Networks} (2017).
\arxiv{1707.00028} 

\bibitem{bib:klimek}
M. D. Klimek, M. Perelstein, 
{\em Neural Network-Based Approach 
to Phase Space Integration}, 
SciPost Phys. \textbf{9} (2020) 053.
\doi{10.21468/SciPostPhys.9.4.053}

\bibitem{bib:mlcwp}
S. Gleyzer, P. Seyfert, S. Schramm (eds.) et al.,
{\em Machine Learning in High Energy Physics 
Community White Paper} (2019).
\arxiv{1807.02876}

\bibitem{bib:bothmann2020}
E. Bothmann et al.,
{\em Exploring phase space 
with Neural Importance Sampling},
SciPost Phys. \textbf{8} (2020) 069.
\doi{10.21468/SciPostPhys.8.4.069}

\bibitem{bib:ml4jets2020}
Generative\,Models\,session,\,ML4Jets2020\,workshop,\,NYU.
\url{https://indico.cern.ch/event/809820/sessions/329213}

\bibitem{bib:butter2019}
A. Butter, T. Plehn, R. Winterhalder,
{\em How to GAN LHC Events},
SciPost Phys. \textbf{7} (2019) 075.
\doi{10.21468/SciPostPhys.7.6.075}

\bibitem{bib:oasis}
K. T. Matchev, P. Shyamsundar,
{\em OASIS: Optimal Analysis-Specific 
Importance Sampling for event generation} (2020).
\arxiv{2006.16972}

\bibitem{bib:biasing}
S. Alioli et al.,
{\em Jet pair production in POWHEG}, 
JHEP04(2011)081.
\doi{10.1007/JHEP04(2011)081}

\bibitem{bib:biasingmg1}
MadGraph, 
Biasing the generation of unweighted partonic events at LO,
\url{https://cp3.irmp.ucl.ac.be/projects/madgraph/wiki/LOEventGenerationBias}.

\bibitem{bib:biasingmg2}
R. Frederix et al., 
{\em Heavy-quark mass effects 
in Higgs plus jets production}, 
JHEP08(2016)006.
\doi{10.1007/JHEP08(2016)006}

\bibitem{bib:sherpaenhanced}
Sherpa Enhance\_Function,
Sherpa 2.0.0 Manual (2013).
\url{https://sherpa.hepforge.org/doc/SHERPA-MC-2.0.0.html\#Enhance\_005fFunction}

\bibitem{bib:biasingpythia}
Pythia8 Sample Main Programs,
\url{http://home.thep.lu.se/~torbjorn/pythia81html/SampleMainPrograms.html}.


\bibitem{bib:alwall09}
J. Alwall, S. de Visscher, F. Maltoni,
{\em QCD radiation in the production 
of heavy colored particles at the LHC}, 
JHEP02(2009)017. 
\doi{10.1088/1126-6708/2009/02/017}

\bibitem{bib:ATL-PHYS-PUB-2017-006}
ATLAS Collaboration,
{\em ATLAS simulation of boson 
plus jets processes in Run 2}, 
ATL-PHYS-PUB-2017-006.
\url{http://cds.cern.ch/record/2261937}


\bibitem{bib:cr1}
S. Gieseke, C. R\"ohr, A. Si\'odmok, 
{\em Colour reconnections in Herwig++},
Eur. Phys. J. C \textbf{72} (2012) 2225.
\doi{10.1140/epjc/s10052-012-2225-5}

\bibitem{bib:cr2}
S. Gieseke et al.,
{\em Colour reconnection 
from soft gluon evolution}, 
JHEP11(2018)149. 
\doi{10.1007/JHEP11(2018)149}

\bibitem{bib:gainer2014}
J. S. Gainer et al.,
{\em Exploring theory space with
Monte Carlo reweighting}, 
JHEP10(2014)78.
\doi{10.1007/JHEP10(2014)078}

\bibitem{bib:mrenna}
S. Mrenna, P. Skands,
{\em Automated parton-shower 
variations in pythia 8},
Phys. Rev. D \textbf{94} (2016) 074005.
\doi{10.1103/PhysRevD.94.074005}

\bibitem{bib:mattelaer2016}
O. Mattelaer, 
{\em On the maximal use of Monte Carlo samples:
re-weighting events at NLO accuracy}, 
Eur. Phys. J. C \textbf{76} (2016) 674. 
\doi{10.1140/epjc/s10052-016-4533-7}

\bibitem{bib:bothmann16}
E. Bothmann, M. Sch\"onherr, S. Schumann,
{\em Reweighting QCD matrix-element 
and parton-shower calculations},
Eur. Phys. J. C \textbf{76} (2016) 590.
\doi{10.1140/epjc/s10052-016-4430-0}

\bibitem{bib:LH17}
J. Bendavid et al., 
{\em Les Houches 2017: Physics at TeV Colliders 
Standard Model Working Group Report},
Proc. Les Houches 2017.
\arxiv{1803.07977}

\bibitem{bib:valassiwdr}
A. Valassi,
{\em Optimising HEP parameter fits 
via Monte Carlo weight derivative regression},
Proc. CHEP2019, Adelaide,
EPJ Web of Conf. \textbf{245} 06038 (2020).
\doi{10.1051/epjconf/202024506038}

\bibitem{bib:hwreweight1}
J. Bellm et al.,
{\em Reweighting parton showers},
Phys. Rev. D \textbf{94} (2016) 034028.
\doi{10.1103/PhysRevD.94.034028}

\bibitem{bib:hwreweight2}
J. Bellm et al.,
{\em Parton-shower uncertainties with Herwig 7: 
benchmarks at leading order},
Eur. Phys. J. C \textbf{76} (2016) 665. 
\doi{10.1140/epjc/s10052-016-4506-x}

\bibitem{bib:atlasreweight}
ATLAS collaboration,
{\em Measurements of WH and ZH production 
in the $H\rightarrow b\bar{b}$ decay channel 
in pp collisions at 13 TeV 
with the ATLAS detector},
ATLAS-CONF-2020-006 (2020).
\url{http://cdsweb.cern.ch/record/2714885}

\bibitem{bib:cranmer2015}
K. Cranmer, J. Pavez, G. Louppe,
{\em Approximating Likelihood Ratios 
with Calibrated Discriminative Classifiers} (2015).
\arxiv{1506.02169}

\bibitem{bib:andreassen}
A. Andreassen, B. Nachman,
{\em Neural Networks 
for Full Phase-space Reweighting 
and Parameter Tuning}, 
Phys. Rev. D \textbf{101} (2020) 091901.
\doi{10.1103/PhysRevD.101.091901} 

\bibitem{bib:peyre}
G. Peyr\'e, M. Cuturi,
{\em Computational Optimal Transport}
Foundations and Trends in Machine Learning \textbf{11} (2019) 355.
\arxiv{1803.00567}

\bibitem{bib:rogoz}
A. Rogozhnikov,
{\em Reweighting with Boosted Decision Trees} (2015).
\url{http://arogozhnikov.github.io/2015/10/09/gradient-boosted-reweighter.html}

\bibitem{bib:frederix2020}
R. Frederix, S. Frixione, S. Prestel. P. Torrielli, 
{\em On the reduction of negative weights 
in MC@NLO-type matching procedures}, 
JHEP07(2020)238.
\doi{10.1007/JHEP07(2020)238}

\bibitem{bib:un2lops}
S. H\"oche, Y. Li, S. Prestel,
{\em Drell-Yan lepton pair production 
at NNLO QCD with parton showers},
Phys. Rev. D \textbf{91} (2015) 074015.
\doi{10.1103/PhysRevD.91.074015}


\bibitem{bib:triple}
S. H\"oche, S. Prestel,
{\em Triple collinear emissions in parton showers},
Phys. Rev. D \textbf{96} (2017) 074017.
\doi{10.1103/PhysRevD.96.074017}

\bibitem{bib:platzer2018}
S. Pl\"atzer, M. Sjodahl, J. Thor\'en, 
{\em Color matrix element corrections 
for parton showers},
JHEP11(2018)009.
\doi{10.1007/JHEP11(2018)009}

\bibitem{bib:martinez2018}
R. \'A. Mart\'inez et al.,
{\em Soft gluon evolution 
and non-global logarithms},
JHEP05(2018)044.
\doi{10.1007/JHEP05(2018)044}

\bibitem{bib:olsson2019}
J. Olsson, S. Pl\"atzer, M. Sjodahl,
{\em Resampling Algorithms for 
High Energy Physics Simulations},
Eur. Phys. J. C \textbf{80} (2020) 934.
\doi{10.1140/epjc/s10052-020-08500-y}
 	
\bibitem{bib:sir}
D. B. Rubin,
{\em A noniterative sampling/importance 
resampling alternative to the data 
augmentation algorithm for creating 
a few imputations when the fraction of 
missing information is modest: 
the SIR algorithm
(comment on an article 
by Tanner and Wong)},
J. Am. Statist. Assoc. 
\textbf{82} (1987) 543.
\doi{10.2307/2289460} 

\bibitem{bib:posresampler}
J. R. Andersen, C. Gutschow, A. Maier, S. Prestel,
{\em A Positive Resampler for Monte Carlo 
Events with Negative Weights},
Eur. Phys. J. C \textbf{80} (2020) 1007.
\doi{10.1140/epjc/s10052-020-08548-w}

\bibitem{bib:nnresampler}
B. Nachman, J. Thaler,
{\em A Neural Resampler for Monte Carlo 
Reweighting with Preserved Uncertainties},
Phys. Rev. D \textbf{102} (2020) 076004.
\doi{10.1103/PhysRevD.102.076004}

\bibitem{bib:hs06}
M. Michelotto et al.,
{\em A comparison of HEP code 
with SPEC benchmarks on multi-core worker nodes},
Proc. CHEP2009, Prague,
J. Phys. Conf. Ser. \textbf{219} (2010) 052009.
\doi{10.1088/1742-6596/219/5/052009}

\bibitem{bib:bmk2020}
A. Valassi et al.,
{\em Using HEP experiment workflows 
for the benchmarking and accounting 
of WLCG computing resources},
Proc. CHEP2019, Adelaide,
EPJ Web of Conf. \textbf{245} 07035 (2020).
\doi{10.1051/epjconf/202024507035}

\bibitem{bib:hoeche2012}
S. H\"oche et al.,
{\em A critical appraisal 
of NLO+PS matching methods},
JHEP09(2012)049. 
\doi{10.1007/JHEP09(2012)049}

\bibitem{bib:hoeche2013}
S. H\"oche et al.,
{\em QCD matrix elements + parton showers: 
the NLO case},
JHEP04(2013)027. 
\doi{10.1007/JHEP04(2013)027}

\bibitem{bib:konstantinov}
D. Konstantinov, 
{\em Optimization of Pythia8}, 
EP-SFT group meeting, CERN (2020). 
\url{https://indico.cern.ch/event/890670}

\bibitem{bib:martin}
T. Martin, 
{\em Computational bottlenecks}, 
ECHEP/ \mbox{Excalibur} Workshop (2020). 
\url{https://indico.cern.ch/event/928965/contributions/3933234}

\bibitem{bib:snowmass2013}
C. Bauer et al., 
{\em Computing for perturbative QCD: 
a Snowmass White Paper} 
SLAC-PUB-15740 (2013).
\arxiv{1309.3598}

\bibitem{bib:lhcctalk}
A. Valassi, E. Yazgan, J. McFayden,
{\em Monte Carlo generators challenges 
and strategy towards HL-LHC},
WLCG meeting with LHCC referees (2020).
\doi{10.5281/zenodo.4028834}

\bibitem{bib:geant4}
S. Agostinelli et al.,
{\em Geant4 — a simulation toolkit},
Nucl. Instr. Meth. A \textbf{506} (2003) 250.
\doi{10.1016/S0168-9002(03)01368-8}

\bibitem{bib:seiskari}
O. Seiskari, J. Kommeri, T. Niemi,
{\em GPU in Physics Computation: 
Case Geant4 Navigation} (2012).
\arxiv{1209.5235}

\bibitem{bib:g4cu}
K. Murakami et al.,
{\em Geant4 based simulation 
of radiation dosimetry in CUDA},
Proc. IEEE NSS/MIC 2013, Seoul.
\doi{10.1109/NSSMIC.2013.6829452}

\bibitem{bib:geantvreport}
G. Corti et al.,
{\em HEP software community meeting 
on GeantV R\&D Panel Report} (2016).
\href{https://hepsoftwarefoundation.org/assets/GeantVPanelReport20161107.pdf}{https://hepsoftwarefoundation.org/assets/GeantVPanelRe\break port20161107.pdf}

\bibitem{bib:canal2019}
P. Canal,
{\em Geant Exascale Pilot Project},
Geant4 R\&D Meeting, CERN (2019).
\url{https://indico.cern.ch/event/809393/contributions/3441114}

\bibitem{bib:gheata2019}
A. Gheata,
{\em Design, implementation and performance 
results of the GeantV prototype},
Outcome of the GeantV prototype 
HSF meeting, CERN (2019).
\url{https://indico.cern.ch/event/818702/contributions/3559124}

\bibitem{bib:geantv}
G. Amadio et al.,
{\em GeantV: results from the prototype 
of concurrent vector particle 
transport simulation in HEP} (2020).
\arxiv{2005.00949}

\bibitem{bib:summit}
Oak Ridge Leadership Computing Facility,
{\em Summit}.
\url{https://www.olcf.ornl.gov/summit}

\bibitem{bib:summit95}
M. Feldman,
{\em New GPU-Accelerated Supercomputers 
Change the Balance of Power on the TOP500},
Top500 news (2018).
\url{https://www.top500.org/news/new-gpu-accelerated-supercomputers-change-the-balance-of-power-on-the-top500}

\bibitem{bib:heget1}
K. Hagiwara et al., 
{\em Fast calculation of HELAS amplitudes 
using graphics processing unit (GPU)},
Eur. Phys. J. C \textbf{66} (2010) 477.
\doi{10.1140/epjc/s10052-010-1276-8}

\bibitem{bib:heget2}
K. Hagiwara et al., 
{\em Calculation of HELAS amplitudes for QCD 
processes using graphics processing unit (GPU)},
Eur. Phys. J. C \textbf{70} (2010) 513.
\doi{10.1140/epjc/s10052-010-1465-5}

\bibitem{bib:heget3}
K. Hagiwara et al., 
{\em Fast computation of MadGraph amplitudes 
on graphics processing unit (GPU)},
Eur. Phys. J. C \textbf{73} (2013) 2608.
\doi{10.1140/epjc/s10052-013-2608-2}

\bibitem{bib:aloha}
P. de Aquino, W. Link, F. Maltoni,
O. Mattelaer, T. Stelzer,
{\em ALOHA: Automatic libraries of helicity amplitudes 
for Feynman diagram computations},
Comp. Phys. Comm. \textbf{183} (2012) 2254.
\doi{10.1016/j.cpc.2012.05.004}

\bibitem{bib:helas1}
H. Murayama, I. Watanabe, K. Hagiwara, 
{\em HELAS: HELicity Amplitude Subroutines 
for Feynman Diagram Evaluations},
KEK-Report 91-11 (1992).
\url{https://lib-extopc.kek.jp/preprints/PDF/1991/9124/9124011.pdf}

\bibitem{bib:helas2}
I. Watanabe, H. Murayama, K. Hagiwara, 
{\em Evaluating Cross Sections 
at TeV Energy Scale by HELAS},
KEK preprint 92-39 (1992).
\url{https://lib-extopc.kek.jp/preprints/PDF/1992/9227/9227039.pdf}

\bibitem{bib:kanzaki1}
J. Kanzaki, 
{\em Monte Carlo integration on GPU},
Eur. Phys. J. C \textbf{71} (2011) 1559. 
\doi{10.1140/epjc/s10052-011-1559-8}

\bibitem{bib:kanzaki2}
J. Kanzaki, 
{\em Application of graphics processing unit (GPU) 
to software in elementary particle/high 
energy physics field},
Procedia Computer Science \textbf{4} (2011) 869.
\doi{10.1016/j.procs.2011.04.092}

\bibitem{bib:roiser20}
S. Roiser,
{\em Progress on porting MadGraph5\_aMC@NLO to GPUs},
HSF/WLCG Virtual Workshop (2020).
\url{https://indico.cern.ch/event/941278/contributions/4101793}

\bibitem{bib:alpaka}
E. Zenker et al.,
{\em Alpaka -- 
An Abstraction Library for Parallel Kernel Acceleration},
Proc. IEEE IPDPSW 2016, Chicago.
\doi{10.1109/IPDPSW.2016.50}

\bibitem{bib:alpakagit}
Alpaka -- Abstraction Library for Parallel Kernel Acceleration.
\url{https://github.com/alpaka-group/alpaka}

\bibitem{bib:oneapi}
Intel oneAPI Toolkits (Beta).
\url{https://software.intel.com/en-us/oneapi}

\bibitem{bib:zmc}
H.-Z. Wu, J.-J. Zhang, L.-G. Pang, Q. Wang,
{\em ZMCintegral: a Package 
for Multi-Dimensional Monte Carlo 
Integration on Multi-GPUs},
Comp. Phys. Comm. \textbf{248} (2019) 106962.
\doi{10.1016/j.cpc.2019.106962}

\bibitem{bib:vegasflow}
S. Carrazza, J. M. Cruz-Martinez,
{\em VegasFlow: accelerating 
Monte Carlo simulation across 
multiple hardware platforms},
Comp. Phys. Comm. \textbf{254} (2020) 107376.
\doi{10.1016/j.cpc.2020.107376}

\bibitem{bib:pdfflow}
S. Carrazza, J. M. Cruz-Martinez, M. Rossi,
{\em PDFFlow: parton distribution 
functions on GPU} (2020).
\arxiv{2009.06635}

\bibitem{bib:gpurng}
S. Y. Jun et al.,
{\em Vectorization of random number generation 
and reproducibility of concurrent 
particle transport simulation},
Proc. ACAT2019, Saas Fee.
\url{https://inspirehep.net/literature/1754423}


\bibitem{bib:evtgenmk}
M. Kreps,
{\em EvtGen status and plans},
in Ref.~\cite{bib:hsfgen2018}.
\url{https://indico.cern.ch/event/751693/contributions/3182956}

\bibitem{bib:athena}
P. Calafiura et al.,
{\em The Athena Control Framework in Production, 
New Developments and Lessons Learned},
Proc. CHEP2004, Interlaken.
\doi{10.5170/CERN-2005-002.456}

\bibitem{bib:athenamp1}
P. Calafiura et al.,
{\em Running ATLAS workloads within 
massively parallel distributed applications 
using Athena Multi-Process framework (AthenaMP)},
Proc. CHEP2015, Okinawa,
J. Phys. Conf. Ser. \textbf{664} (2015) 072050.
\doi{10.1088/1742-6596/664/7/072050}

\bibitem{bib:athenamp2}
J. Elmsheuser et al.,
{\em ATLAS Grid Workflow Performance Optimization}, 
Proc. CHEP2018, Sofia, 
EPJ Web of Conf. \textbf{214}, 03021 (2019). 
\doi{10.1051/epjconf/201921403021}

\bibitem{bib:athenamt}
C. Leggett et al.,
{\em AthenaMT: upgrading the ATLAS software framework 
for the many-core world with multi-threading},
Proc. CHEP2016, San Francisco,
J. Phys. Conf. Ser. \textbf{898} (2017) 042009.
\doi{10.1088/1742-6596/898/4/042009}

\bibitem{bib:atlassim}
M. Bandieramonte et al.,
{\em Multi-threaded simulation for ATLAS: 
challenges and validation strategy},
Proc. CHEP2019, Adelaide,
EPJ Web of Conf. \textbf{245} 02001 (2020).
\doi{10.1051/epjconf/202024502001}

\bibitem{bib:geant4mt}
J. Allison et al.,
{\em Recent developments in Geant4},
Nucl. Instr. Meth. A \textbf{835} (2016) 186.
\doi{10.1016/j.nima.2016.06.125}

\bibitem{bib:gauss}
M. Clemencic et al.,
{\em The LHCb Simulation Application, 
Gauss: Design, Evolution and Experience},
Proc. CHEP2010, Taipei,
J. Phys. Conf. Ser. \textbf{331} (2011) 032023.
\doi{10.1088/1742-6596/331/3/032023}

\bibitem{bib:gaudi1}
G. Barrand et al.,
{\em GAUDI -- A software architecture and framework 
for building HEP data processing applications},
Comp. Phys. Comm. \textbf{140} (2001) 45.
\doi{10.1016/S0010-4655(01)00254-5}

\bibitem{bib:gaudi2}
M. Clemencic et al.,
{\em Recent developments in the 
LHCb software framework Gaudi},
Proc. CHEP2009, Prague,
J. Phys. Conf. Ser. \textbf{219} (2010) 042006.
\doi{10.1088/1742-6596/219/4/042006}

\bibitem{bib:gaussmp}
F. Stagni, A. Valassi, V. Romanovskiy,
{\em Integrating LHCb workflows 
on HPC resources: status and strategies},
Proc. CHEP2019, Adelaide,
EPJ Web of Conf. \textbf{245} 09002 (2020).
\doi{10.1051/epjconf/202024509002}

\bibitem{bib:gaussino}
B. G. Siddi, D. M\"uller,
{\em Gaussino - a Gaudi-Based 
Core Simulation Framework},
Proc. IEEE NSS/MIC 2019, Manchester.
\doi{10.1109/NSS/MIC42101.2019.9060074}

\bibitem{bib:cmssw0}
E. Sexton-Kennedy, P. Gartung, C. D. Jones, D. Lange,
{\em Implementation of a Multi-threaded Framework 
for Large- scale Scientific Applications},
Proc. ACAT2014, Prague,
J. Phys. Conf. Ser. \textbf{608} (2015) 012034.
\doi{10.1088/1742-6596/608/1/012034}

\bibitem{bib:cmssw1}
C. D. Jones at al.,
{\em Using the CMS Threaded Framework 
In A Production Environment},
Proc. CHEP2015, Okinawa,
J. Phys. Conf. Ser. \textbf{664} (2015) 072026.
\doi{10.1088/1742-6596/664/7/072026}

\bibitem{bib:cmssw2}
C. D. Jones,
{\em CMS event processing 
multi-core efficiency status},
Proc. CHEP2016, San Francisco,
J. Phys. Conf. Ser. \textbf{898} (2017) 042008.
\doi{10.1088/1742-6596/898/4/042008}

\bibitem{bib:cmsmem}
M. Hildreth, V. N. Ivanchenko, D. J. Lange,
{\em Upgrades for the CMS simulation},
Proc. CHEP2016, San Francisco,
J. Phys. Conf. Ser. \textbf{898} (2017) 042040.
\doi{10.1088/1742-6596/898/4/042040}

\bibitem{bib:bendavidcms}
J. Bendavid,
{\em CMS experience with current generators},
Argonne and Fermilab Workshop 
on Beyond Leading Order Calculations on HPCs,
Fermilab (2016).
\url{https://indico.cern.ch/event/557731/contributions/2309458}

\bibitem{bib:cmstwiki}
C. Li, 
{\em The CMS Offline WorkBook:
multithreading in generators},
CMS Public Web (2021).
\url{https://twiki.cern.ch/twiki/bin/view/CMSPublic/WorkBookGenMultithread}


\bibitem{bib:lha01}
E. Boos et al.,
{\em Generic User Process Interface 
for Event Generators},
Proc. Physics at TeV colliders Workshop, 
Les Houches (2001).
\arxiv{hep-ph/0109068}

\bibitem{bib:lhef}
J. Alwall et al.,
{\em A standard format for Les Houches Event Files},
Comp. Phys. Comm. \textbf{176} (2007) 300. 
\doi{10.1016/j.cpc.2006.11.010}

\bibitem{bib:mpi}
J. J. Dongarra et al.,
{\em A message passing standard 
for MPP and workstations},
Comm. ACM (1996).
\doi{10.1145/233977.234000}

\bibitem{bib:hoeche2019}
S. H\"oche, S. Prestel, H. Schulz, 
{\em Simulation of vector boson
plus many final jets 
at the high luminosity LHC}, 
Phys. Rev. D \textbf{100} (2019) 0140124. 
\doi{10.1103/PhysRevD.100.014024}

\bibitem{bib:cori}
NERSC, {\em Cori}.
\url{https://docs.nersc.gov/systems/cori}

\bibitem{bib:childers2017}
J. T. Childers et al.,
{\em Adapting the serial Alpgen 
parton-interaction generator 
to simulate LHC collisions 
on millions of parallel threads},
Comp. Phys. Comm. ~\textbf{210} (2017) 54.
\doi{10.1016/j.cpc.2016.09.013}

\bibitem{bib:mira}
Argonne Leadership Computing Facility,
{\em Mira}.
\url{https://www.alcf.anl.gov/alcf-resources/mira}


\bibitem{bib:mattelaer2019}
O. Mattelaer, 
{\em MG5aMC status and plans},
in Ref.~\cite{bib:hsfgen2018}.
\url{https://indico.cern.ch/event/751693/contributions/3182951}

\bibitem{bib:mcfm}
J. Campbell, T. Neumann,
{\em Precision phenomenology with MCFM},
JHEP12(2019)034.
\doi{10.1007/JHEP12(2019)034}

\bibitem{bib:omp}
The OpenMP API specification 
for parallel programming,
\url{https://www.openmp.org}

\bibitem{bib:kiran}
O. Mattelaer, K. Ostrolenk, 
{\em Speeding up MadGraph5\_aMC@NLO},
MCNET-21-01 (2021).
\arxiv{2102.00773}


\bibitem{bib:minlo_nnlops1}
K. Hamilton et al.,
{\em Merging H/W/Z + 0 and 1 jet 
at NLO with no merging scale: 
a path to parton shower + NNLO matching},
JHEP05(2013)082. 
\doi{10.1007/JHEP05(2013)082}


\bibitem{bib:geneva2}
S. Alioli et al.,
{\em Matching fully differential NNLO 
calculations and parton showers},
JHEP06(2014)089.
\doi{10.1007/JHEP06(2014)089}

\bibitem{bib:minnlops}
P. F. Monni et al.,
{\em MiNNLOPS: A new method to match 
NNLO QCD to parton showers},
JHEP05(2020)143.
\doi{10.1007/JHEP05(2020)143}

\bibitem{bib:mazzitelli}
J. Mazzitelli et al.,
{\em Next-to-next-to-leading order 
event generation 
for top-quark pair production},
CERN-TH-2020-219 (2020).
\arxiv{2012.14267}

\bibitem{bib:minlo_nnlops2}
K. Hamilton et al.,
{\em NNLOPS simulation of Higgs boson production},
JHEP10(2013)222. 
\doi{10.1007/JHEP10(2013)222}

\bibitem{bib:chawdry}
H. A. Chawdhry et al.,
{\em NNLO QCD corrections to three-photon 
production at the LHC},
JHEP02(2020)57. 
\doi{10.1007/JHEP02(2020)057}

\bibitem{bib:czakon2016}
M. Czakon, P. Fiedler, D. Heymes, A. Mitov,
{\em NNLO QCD predictions for fully-differential 
top-quark pair production at the Tevatron},
JHEP05(2016)034.
\doi{10.1007/JHEP05(2016)034}

\bibitem{bib:grazzini2019}
S. Catani et al.,
{\em Top-quark pair production at the LHC: 
fully differential QCD predictions at NNLO},
JHEP07(2019)100. 
\doi{10.1007/JHEP07(2019)100}

\bibitem{bib:dire}
S. H\"oche, S. Prestel, 
{\em The midpoint between 
dipole and parton showers},
Eur. Phys. J. C \textbf{75} (2015) 461. 
\doi{10.1140/epjc/s10052-015-3684-2}

\bibitem{bib:vincia}
N. Fischer, S. Prestel, M. Ritzmann, P. Skands,
{\em VINCIA for hadron colliders},
Eur. Phys. J. C \textbf{76} (2016) 589.
\doi{10.1140/epjc/s10052-016-4429-6}

\bibitem{bib:deductor}
Z. Nagy, D. E. Soper,
{\em Jets and threshold summation in deductor},
Phys. Rev. D \textbf{98} (2018) 014035.
\doi{10.1103/PhysRevD.98.014035}

\bibitem{bib:nlodglap}
S. H\"oche, F. Krauss, S. Prestel, 
{\em Implementing NLO DGLAP evolution 
in parton showers},
JHEP10(2017)093. 
\doi{10.1007/JHEP10(2017)093}

\bibitem{bib:dulat}
F. Dulat, S. H\"oche, S. Prestel, 
{\em Leading-color fully differential 
two-loop soft corrections 
to QCD dipole showers},
Phys. Rev. D \textbf{98} (2018) 074013.
\doi{10.1103/PhysRevD.98.074013}

\bibitem{bib:dasgupta}
M. Dasgupta et al.,
{\em Parton showers beyond 
leading logarithmic accuracy},
Phys. Rev. Lett. \textbf{125} (2020) 052002.
\doi{10.1103/PhysRevLett.125.052002}

\bibitem{bib:actis2013}
S. Actis et al.,
{\em Recursive generation of one-loop 
amplitudes in the Standard Model},
JHEP04(2013)037. 
\doi{10.1007/JHEP04(2013)037}

\bibitem{bib:kallweit2015}
S. Kallweit et al.,
{\em NLO electroweak automation 
and precise predictions for 
W+multijet production at the LHC},
JHEP04(2015)012.
\doi{10.1007/JHEP04(2015)012}

\bibitem{bib:schonherr2018}
M. Sch\"onherr, 
{\em An automated subtraction 
of NLO EW infrared divergences},
Eur. Phys. J. C \textbf{78} (2018) 119. 
\doi{10.1140/epjc/s10052-018-5600-z}

\bibitem{bib:frederix2018}
R. Frederix et al.,
{\em The automation of next-to-leading 
order electroweak calculations},
JHEP07(2018)185. 
\doi{10.1007/JHEP07(2018)185}


\end{thebibliography}
\end{document}